\documentclass[aps,pre,twocolumn,superscriptaddress,showpacs,longbibliography]{revtex4-1}

\usepackage{amsfonts,amssymb,amsmath,latexsym,epsfig,wasysym} 
\usepackage[sort&compress]{natbib}

\usepackage{color}
\definecolor{bluecolor}{rgb}{0,0.,1.}

\definecolor{redcolor}{rgb}{.7,0.,0.}

{}

\begin{document}

\title{Stochastic model for the vocabulary growth in natural languages}

\author{Martin Gerlach} 
\affiliation{Max Planck Institute for the Physics of Complex Systems, 01187 Dresden, Germany}
\author{Eduardo G. Altmann} 
\affiliation{Max Planck Institute for the Physics of Complex Systems, 01187 Dresden, Germany}

\begin{abstract}
We propose a stochastic model for the number of different words in a given database which incorporates the dependence on the database
size and historical changes. The main feature of our model is the existence of two different classes of words: (i) a finite number of
core-words which have higher frequency and do not affect the probability of a new word to be used; and (ii) the remaining virtually infinite
number of noncore-words which have lower frequency and once used reduce the probability of a new word to be used in the future. 
Our model relies on a careful analysis of the google-ngram database of books published in the last centuries and its main consequence is the generalization of Zipf's and Heaps' law to two scaling regimes. 
We confirm that these generalizations yield the best simple description of the data among generic descriptive models and that the two free parameters depend only on the language but not on the database. 
From the point of view of our model the main change on historical time scales is the composition of the specific words included in the finite list of core-words, which we observe to decay exponentially in time with a rate of approximately 30 words per year for English. 
\end{abstract}

\pacs{89.65.-s, 89.75.Da, 87.23.Ge, 05.10.-a}
\maketitle


\section{Introduction}\label{sec.intro}
Even in our time of big data~\cite{Michel2011:0, Gao2012:0, Petersen2012:0} there is no indication of a saturation of the vocabulary size (total number of different words) with increasing database size.
In order to clarify whether it is meaningful to estimate a vocabulary size in the limit of infinitely large databases, it is essential to understand not only the birth and death of words~\cite{Pagel2007:0, Lieberman2007:0, Levary2012:0}, but
also the process governing the usage of new words and its dependence on database size.
The interest in this problem is motivated by fundamental linguistic studies~\cite{Wimmer1999:0,Baayen2001} as well as by applications in search engines, which require an estimation of the number of different words in a given database~\cite{Baeza-Yates2000:0, Williams2005:0, Croft2009}.

The scaling between the number of different words, $N$, and the size of the database in words, $M$, as $N \sim M^\lambda$ is known as Heaps' law~\cite{Heaps1978} and has been studied in different linguistic~\cite{Serrano2009:0,Bernhardsson2009:0,Sano2012:0} and non-linguistic~\cite{Cattuto2009:0,Benz2008:0} contexts. 
The universality and interest of this empirical scaling is surpassed only by Zipf's law~\cite{Zipf1936}, which states that the frequency $F(r)$ of the $r$-th most frequent word in a database decays as~$F(r) \sim 1/{r}$. 
The relation between Heaps' and Zipf's law has been the subject of great recent interest~\cite{Leijenhorst2005:0,Zanette2005:0,Eliazar2011:1}. Furthermore, it is well known that deviations of the Heaps'- and Zipf's-laws are observed in the tails of Heaps'- and Zipf's- plots (i.e., for large $N$ and $r$, respectively)~\cite{Montemurro2001:1,Li2010:0,JAGER2012:0}.
Similar deviations of fat-tailed distributions appear in a variety of social and physical systems~\cite{Newman2005:0, Stumpf2012:0} and are crucial when extrapolating to the limit of large databases.

In this paper we propose a stochastic growth model whose predictions go beyond the simpler scalings of Heaps' and Zipf's law and are
compatible with actual observations in the tail of the corresponding distributions. 
Our model is in the same spirit of, but differs from, the simpler versions of Yule's-, Simon's-, Gibrat's-, and preferential attachment- growth models~\cite{Yule1925:0,Mitzenmacher2004:0, Newman2005:0,Simkin2010:0}, because it contains two categories of words and leads to two scaling regimes in the Heaps'- and Zipf's- plots.
These findings are supported by a statistical analysis of the google-ngram database indicating that the only two free parameters needed in the description of these scalings remain unchanged over centuries, depend only on the language, and that there is a slow change of words belonging to each category.
The latter adds to the recent interest in language dynamics as a complex system~\cite{Castellano2009:0,BARONCHELLI2012:0}.

The paper is organized as follows: in Sec.~\ref{sec.emp} we present statistical analysis of the
google-ngram database in terms of word frequencies  as well as the growth of the vocabulary.
This will then lead us to the formulation of our stochastic model for the vocabulary growth in Sec.~\ref{sec.model}. 
In Sec.~\ref{sec.hist} we investigate dynamical aspects on historical time scales within the framework of our model.

\section{Data Analysis}\label{sec.emp}
\begin{figure*}[!bt]
\centering
\includegraphics[width=2\columnwidth]{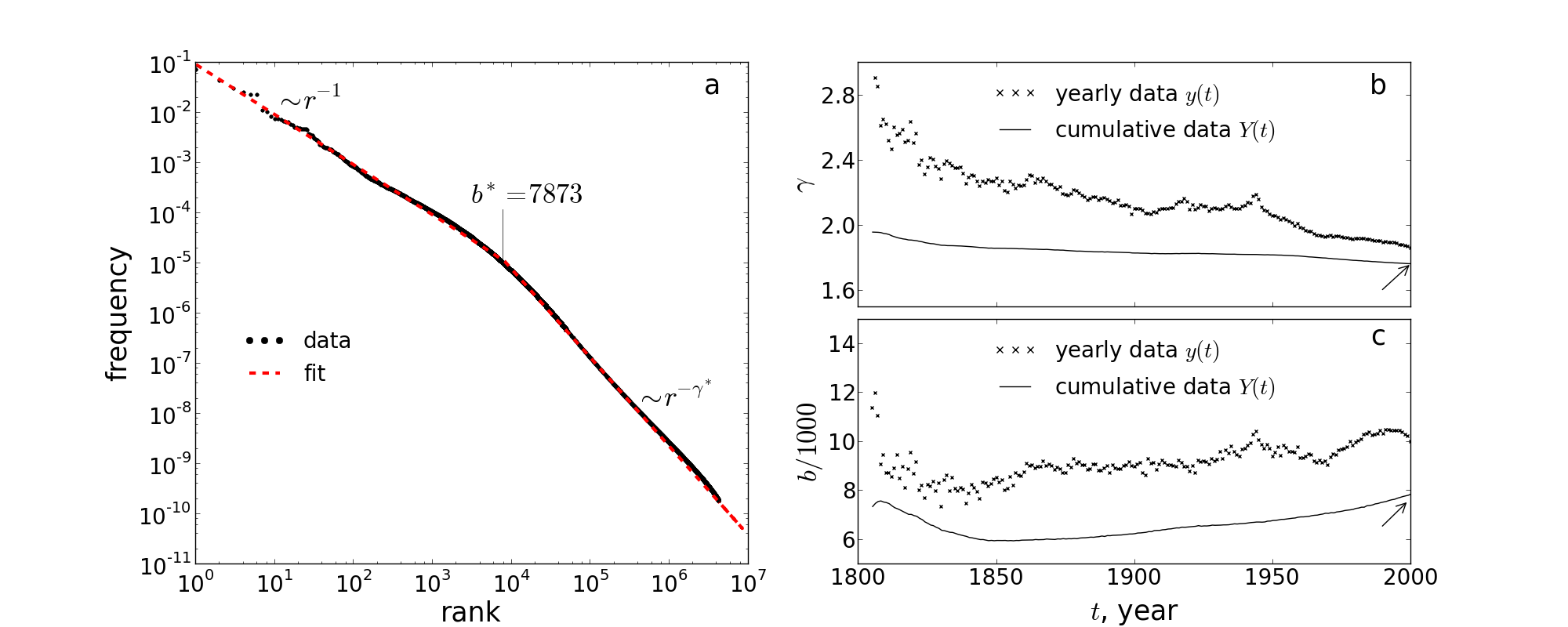}
\caption{Rank-frequency distribution shows double scaling behavior (Zipf's plot). a) Rank-frequency distribution for the English
  database $Y(2000)$ (solid) and a ML-fit of Eq.~(\ref{eq.modeldp}) (dashed).  b+c) parameters $\gamma$ and $b$ obtained from ML-fits of
  Eq.~(\ref{eq.modeldp}) to yearly $y(t)$ (x-symbols) and accumulated $Y(t)$  (solid) database. Arrows indicate the values of the parameters
  $\gamma^*$ and $b^*$ obtained for the fit in a). Results are shown for the time range $t\in[1805,200]$ in which data is most reliable,
  accumulation starts in $t_0=1520$. }
\label{fig.Zipfs}
\end{figure*}
\subsection*{Data}

The main motivation for our model comes from empirical observations. 
As databases, we use the google-ngram corpus~\cite{Michel2011:0} for English, German, French, Spanish, and Russian, which provide data of the word-frequencies (occurring in printed books) with a yearly resolution for a period of several hundred years (1520-2000). 
Our main interest in this database stems from its large size (several millions of books with $>10^{11}$ words) and from the long time span it covers (thus enabling us to trace historical changes in the usage of language).
We consider as words only the $1$-grams consisting uniquely by letters present in the alphabet of the corresponding language.
This pragmatic definition reduces the effect of symbolic sequences, foreign words, numbers, or scanning problems in our observations and should be taken into account when interpreting our findings about the vocabulary.
For each language we use two different partitions of the database: i) yearly ($y$), in which case $y(t)$ corresponds to the database
of the year $t$; and ii) cumulative ($Y$), in which case $Y(t)=\sum_{t'=to}^{t} y(t')$. 
We consider only words which appeared at least $n=41$ times in order to avoid biases due to the filtering mechanism used in the google-ngram database, see Supplemental Material (SM) Sec.~I for further details~\cite{SM}.
Here we show our detailed analysis for the largest database (English, $t_0=1520$, $t \in [1805,2000]$). 
For the other $4$ languages we report the main findings and leave the details for the SM~\cite{SM}.

\subsection*{Zipf's analysis}

Our first empirical analysis focuses on the distribution of word frequencies. 
In his seminal work, Zipf proposed that the frequency of the $r$-th most frequent word in a given text is given by $F(r) = F(1)/r$~\cite{Zipf1936}. 
It is easy to see that this scaling has to break for large $r$:  due to the divergence of the harmonic series, for sufficiently large databases one arrives at $\sum_{r=1}^{N} F(r) > 1$ (sum of frequencies larger than text size). In English $F(1)\approx0.07$ (the frequency of ``the'') and $\sum_{r=1}^N F(r)>1$ for $N \approx 10^6$, meaning that $F(r)$ has to decay faster than $1/r$ for $r\gtrsim10^6$. 
This well-known expectation, which is clearly seen in our data shown in Fig.~\ref{fig.Zipfs}(a), motivated numerous different generalization of Zipf's proposal~\cite{Mandelbrot1953:0,Tuldava1996:0,Baek2011:0}. 
While many of these proposals were shown to provide a better account of particular databases, they remain in a great extent unsatisfactory because they lack the simplicity and universality of Zipf's original proposal (e.g., the parameters vary depending on the size, topic or date of publication of the analyzed texts~\cite{Cohen1997:0,Cancho2005:0}). 
Motivated by the new magnitude of our large database, we apply rigorous statistical tests to determine which of the previously proposed distributions provide a better account of the data. 
We select $7$ of the most popular previously-proposed heavy-tailed distributions with at most $2$ free parameters~\cite{Baayen2001,Li2010:0,JAGER2012:0}: 
power-law, two power-laws, shifted power-law, log-normal, Weibull, and power-laws with exponential cutoffs (in the tail and beginning,
respectively).
The parameters for each distribution were obtained numerically by means of Maximum Likelihood (ML) estimation~\cite{Press2007}. 
In addition we i) calculate the probability that the data was generated by that model ($\chi^2$ $p$-value~\cite{D'Agostino1986,Taylor1997}) and ii) compare which model is more likely to describe the data (relative likelihood~\cite{Akaike1974:0,Burnham2002}) for each fit (for details see SM-Sec.~II A~\cite{SM}).
 
The results show that it is extremely unlikely ($p<10^{-15}$) that the data was drawn exactly from any of the proposed distributions, a consequence of the large databases which makes any small (true) deviation incompatible with these simple fits. 
On the other hand, the results show unequivocally that for English the distribution with two power-laws is the best fit ($1-p<10^{-15}$) for all databases with a size larger than $10^9$ words. We confirm that the double power-law is also the best fit for the English Wikipedia, a strong indication of the validity of this result in databases of different origin (see SM-Sec.~IIB for the detailed analysis on both databases~\cite{SM} which uses methods reported in Refs.~\cite{Methods}).

We now discuss in detail the best two-parameter model we identify from our data:
\begin{equation}\label{eq.modeldp}
 F_{dp}(r;\gamma,b)=C\begin{cases}
r^{-1}, & r\leq b\\
b^{\gamma-1}r^{-\gamma} & r>b,
\end{cases}
\end{equation}
characterizing a double power-law (dp), where $b$, and $\gamma$ are free-parameters, and $C=C(\gamma,b)$ is the normalization
constant~\cite{footnote:dp}.
The effect of the threshold $n$ applied to the frequency of words is that, in practice, data of $F(r)$ is limited to $F(r)\geq n/M$ ($M$ is the observed number of words). 
The original Zipf's law is recovered for high-frequency words and a critical rank $r=b$ determines a transition to a power-law with exponent $\gamma$. 
Double power-laws were proposed as a generalization of Zipf's law in Ref.~\cite{Naranan1998:0} and further investigated in
Refs.~\cite{Cancho2001:1,Petersen2012:1}. 
These insightful works used distributions with two power-law exponents $\gamma_1,\gamma_2$ and were motivated by the visual inspection of double logarithmic plots. 
Our improved statistical analysis confirm and extend these observations for the simpler distribution Eq.~(\ref{eq.modeldp}).
Besides the likelihood analysis and visual inspection given in Fig.~\ref{fig.Zipfs}, a third strong evidence in favor of distribution (\ref{eq.modeldp}) comes from the comparison of the estimated parameters of different corpora shown in Fig.~\ref{fig.Zipfs}(b,c). 
Very similar values $b\in[7\cdot10^3,12\cdot10^3]$ and $\gamma \in [1.8,2.5]$ were obtained for non-overlapping databases (for the English Wikipedia: $b=7830$, $\gamma = 1.68$), and the fluctuations become smaller for increasing database size.
These observations strongly suggest that the same fixed parameters provide a good description of all English texts (e.g., $y(1900)$ and $y(2000)$). 
Therefore, hereafter we do not consider individual fits for each database and instead assume that Eq.~(\ref{eq.modeldp}) is valid with $b=b^*=7873$ and $\gamma=\gamma^*=1.77$, values obtained for our largest database $Y(2000)$.
\linebreak
\indent Similar findings also apply to the other languages. 
In Tab.~\ref{tab.MLres} we summarize the parameters $\gamma^*$ and $b^*$ obtained from a ML-fit of the largest database $Y(2000)$ of the respective language to Eq.~(\ref{eq.modeldp}).
French and Spanish are also best described by Eq.~(\ref{eq.modeldp}) for databases exceeding a particular size and yield values for $\gamma^*$ and $b^*$ similar to English. 
For German and Russian Eq.~(\ref{eq.modeldp}) constitutes only the second best model. 
However, we have strong indications that it provides a better account of the tails ($r\gg b^*$) and therefore we expect that even larger databases will reveal the double power-law as the best fit also in these languages (see SM-Sec.~II B for details~\cite{SM}). 
Apart from being the smallest databases among the investigated languages, another feature affecting the fitting in German and especially in Russian is the higher degree of inflection in the morphology of these languages. 
We recall that no lemmatization was applied in our definition of words and, therefore, inflected words (obtained, e.g., by adding a suffix) are counted as distinct words.
This reasoning explains the higher measured values of $b^*$ (vocabulary in the $r^{-1}$ regime).
From the fitting perspective, however, the large values of $b^*$ in German and Russian require even larger databases to characterize the deviations from the $r^{-1}$ regime for $r\gg b^*$.
\begin{table}[ht]
\begin{tabular*}{0.75\columnwidth}{@{\extracolsep{\fill}}|l||c|c|c|}
\hline
  language\,\,\,\,\,\,\,\,\,\,  & $b^*$ & $\gamma^*$ & $C^* =C(\gamma^*,b^*)$  \\
\hline
\hline
English & $7,873$ & $1.77$ & $0.0922$ \\
\hline
French & $8,208$ & $1.78$ & $0.0920$ \\
\hline
Spanish & $8,757$ & $1.78$ & $0.0915$ \\
\hline
German & $19,863$ & $1.62$ & $0.0828$\\
\hline
Russian & $62,238$ & $1.94$ & $0.0789$\\
\hline
\end{tabular*}
\caption{Parameters $b^*$, $\gamma^*$, and $C^* =C(\gamma^*,b^*)$ obtained from ML-fit of Eq.~(\ref{eq.modeldp}) obtained for the largest database $Y(2000)$ for all considered languages.}
\label{tab.MLres}
\end{table}

\subsection*{Heaps' analysis}
\begin{figure*}[!bt]
\centering
\includegraphics[width=2\columnwidth]{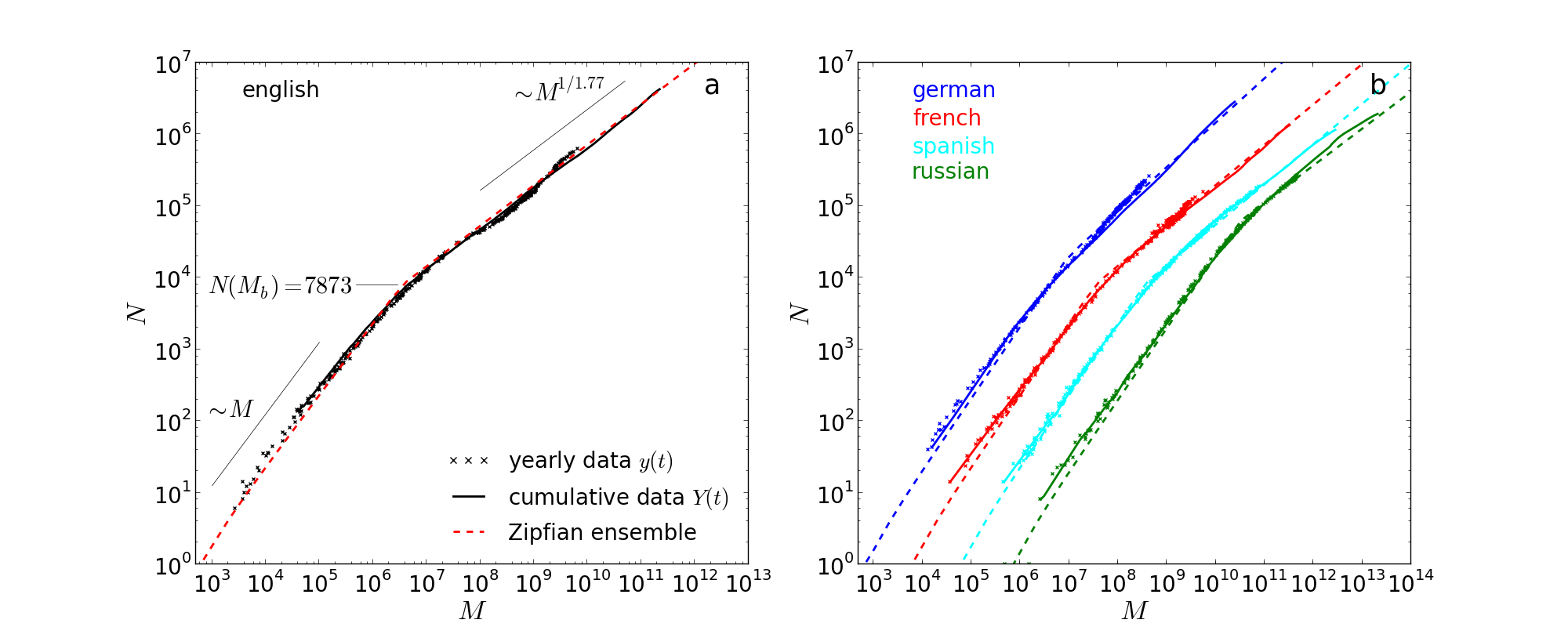}\\
\includegraphics[width=2\columnwidth]{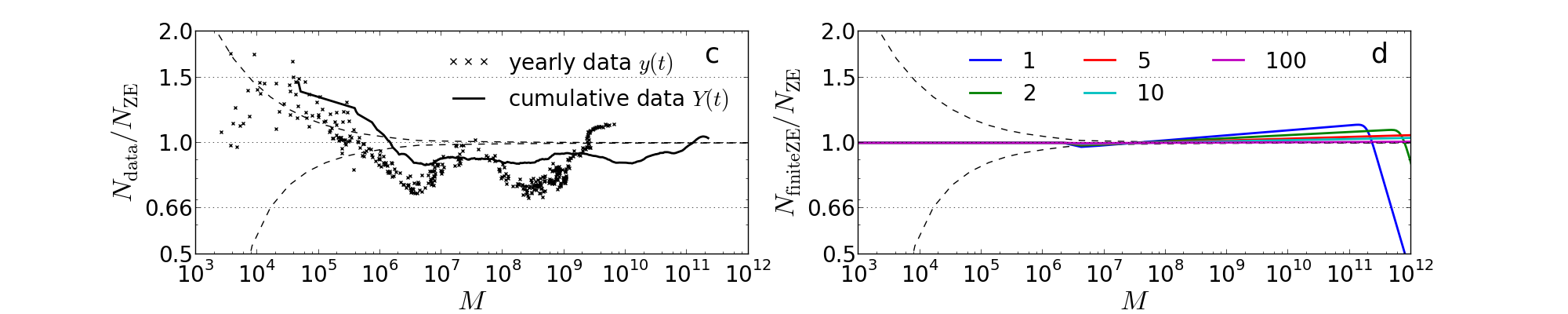}
\caption{Vocabulary $N$ as a function of database size $M$ (Heaps' plot). a) Number of distinct words as a function of the number of words for yearly $y(t)$ (x-symbols) database, cumulative $Y(t)$ (solid) database, and the Zipfian ensemble (dashed) assuming $n=41$ and the rank-frequency distribution Eq.~(\ref{eq.modeldp}) with $b^*=7873$ and $\gamma^* = 1.77$.   b) Same curves as in a) but for different languages showing the same scaling behaviour. In order to increase visibility the curves for French, Spanish, and Russian were shifted, respectively, by one, two, and three decades with respect to their x-values. c) Difference of the curves in a): Deviation of the data $y(t)$ and $Y(t)$ ($N_{\text{data}}$) from the ZE growth curve ($N_{\text{ZE}}$). The dashed lines show the $95\%$-confidence interval of the ZE. d) Deviation of a ZE growth curve with a hypothetically finite vocabulary ($N_{\text{finiteZE}}$) from the ZE growth curve with infinite vocabulary ($N_{\mathrm{ZE}}$) assuming rank-frequency distribution Eq.~(\ref{eq.modeldp}). Possible size of total vocabulary is given in units $k$ of the number of observed distinct words in $Y(2000)$, such that $N_{\mathrm{ZE}}^{\mathrm{max}} = k\cdot4\,263\,717$ with $k=1,2,5,10,100$. Since for $M\longrightarrow \infty:$ $N_{\text{finiteZE}}(M) \longrightarrow N_{\mathrm{ZE}}^{\mathrm{max}}$ the deviation for $k=1$ becomes already large for $M>10^{11}$.
}
\label{fig.Heaps}
\end{figure*}
We now turn to our second empirical analysis: the dependence of the number of different words, $N$, on
the size of the database, i.e. total number of words, $M$. 
The classical result for this relation is the empirical Heaps' law ~\cite{Heaps1978}, which states that $N \sim M^{\lambda}$ with
$\lambda \in [0,1]$ ($A \sim B$ indicates that $A/B=$constant for large $B$).
We start searching for the consequences of our previous observations in the Zipf's analysis to this new problem.
A simple and powerful approach is the so-called Zipfian ensemble (ZE)~\cite{Eliazar2011:1}, which can be traced~\cite{Petersen2012:1}
  back to Mandelbrot~\cite{Mandelbrot1961}. It assumes that the occurrence of every possible word is governed by a Poisson process with an intensity proportional to its frequency (see SM-Sec.~III A~\cite{SM}).
It was shown that under this or similar assumptions (e.g., stochastic processes with fixed frequencies for words), asymptotically Heaps' law can be interpreted as a direct consequence of a Zipfian rank frequency distribution $F(r) \sim r^{-\gamma}$~\cite{Baeza-Yates2000:0, Leijenhorst2005:0, Serrano2009:0, Bernhardsson2009:0, Eliazar2011:1} and vice versa~\cite{Simon1955:0, Zanette2005:0, Masucci2011:0}, where $\gamma = 1/\lambda$~\cite{Mandelbrot1961}.   
Here we want to draw attention to the fact that these observations are not restricted to Zipf's and Heaps' laws, i.e., assuming a stochastic model, the relationship between $F(r)$ and $N(M)$ can always be established.
The expectation of the ZE of Eq.~(\ref{eq.modeldp}) with a threshold $n\gg1$ is (see SM-Sec.~III B~\cite{SM})
\begin{equation}\label{eq.modelH}
 N_{\mathrm{dp}}(M;\gamma,b) = C_n \begin{cases}
M, & M \ll M_b\\
M_b^{1-1/\gamma}M^{1/\gamma}, & M \gg M_b,
\end{cases}
\end{equation}
where $M_b$ is the number of words such that $N(M_b)=b$ and the scaling constant $C_n = C/n$ [$C\approx F(1)$ being the frequency of the most common word, as can be seen from Eq.~(\ref{eq.modeldp})].
Thus, the effect of the threshold $n$ applied to the growth curve of the vocabulary simply amounts to rescaling the constant $C$.
While the expected (average) number of distinct words over many realizations of the stochastic process leads to a sharp transition between the two regimes, the values of  $N_{\mathrm{dp}}(M\approx M_b)$ might depend more strongly on the particular realization.

In Fig.~\ref{fig.Heaps} we show that the data in the google-ngram database obeys the scalings of Eq.~(\ref{eq.modelH}).  
In Fig.~\ref{fig.Heaps}(a) we present the $N(M)$ curve for English. 
While for the yearly database~$y(t)$ we obtain a set of points for each $t$, the cumulative database $Y(t)$ builds a curve of vocabulary growth for increasing $t$.
Despite the differences in these databases, all the data lie in a relatively narrow region of the plot which resembles a single curve compatible with the double scaling of Eq.~(\ref{eq.modelH}). 
This curve is well described by the $N(M)$ curve obtained from the combination of the double power-law distribution Eq.~(\ref{eq.modeldp}) with fixed parameters ($\gamma^*$, $b^*$) and the assumption of Poisson usage of words, in the spirit of the ZE. 
Similar observations apply to all considered languages, as shown in Fig.~\ref{fig.Heaps}(b). 
On closer inspection, Fig.~\ref{fig.Heaps}(c), the fine details of the $N(M)$ curve are not compatible with the fluctuations expected from the strongly simplifying assumptions of the ZE.
Nevertheless, it is remarkable that the agreement between model and data remains within $50\%$ for different databases and over $9$ orders of magnitude in size.

Here it is worth revisiting the question about the finitude of the vocabulary. 
Even after more than $10^6$ different words the $N(M)$ data in Fig.~\ref{fig.Heaps} does not seem to saturate. 
To further investigate this point, we perform the ZE with the same rank-frequency distribution from Eq.~(\ref{eq.modeldp}) (fixed $b^*$, $\gamma^*$) but varying the maximum possible number of different words $N_{\mathrm{ZE}}^{\mathrm{max}}$, e.g., $1$, $2$, $5$ ,$10$, and $100$ times the observed number of distinct words in our largest database $Y(2000)$. 
It can be seen in Fig.~\ref{fig.Heaps}(d) that the differences for the predicted growth curves for such different hypothetical vocabulary sizes are negligible compared to the fluctuations of the real data.
From this we conclude that given the data accessible so far the possible vocabulary can be regarded for all practical purposes to be infinite (although bounded by combinatorial arguments due to a finite alphabet and word length). 
The fact that the same distribution Eq.~(\ref{eq.modeldp}) with fixed parameters accounts for the observation across all years shows that the observation of different number of words is driven mainly by the different database size and not by a change in vocabulary richness over time.

\section{Model}\label{sec.model}

In this section we propose a simple generative model which recovers and allows for an improved interpretation of the double scalings in our empirical findings -- Eqs.~(\ref{eq.modeldp}) and~(\ref{eq.modelH}).

\begin{figure}[t]
\centering
\includegraphics[width=0.97\columnwidth]{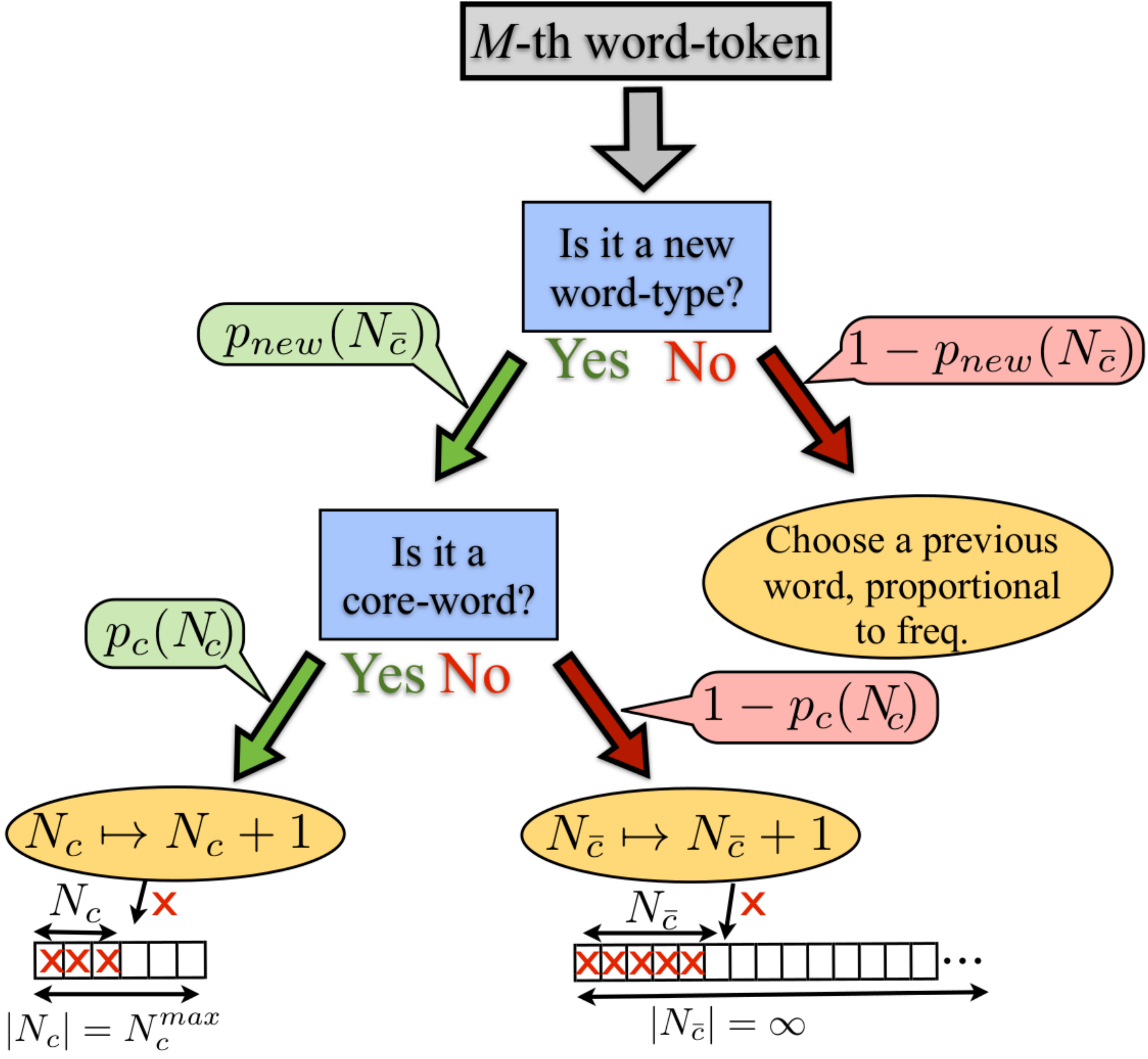}
\caption{Illustration of our generative model for the usage of new words. }
\label{fig.model}
\end{figure}

Our approach is different from Zipf's original explanation based on a principle of least effort between speakers and listeners~\cite{Zipf1936,Murtra2011:0}, but instead is in line with the tradition of Yule-type stochastic growth models explaining fat-tailed distributions~\cite{Yule1925:0,Mitzenmacher2004:0, Newman2005:0,Simkin2010:0}. 
The main novelty in our model is that it contains two classes of word-types: a core vocabulary and a noncore vocabulary~\cite{Cancho2001:1}.
At each step a word (i.e. word-token) is drawn ($M\mapsto M+1$) and attributed to one of the distinct words (i.e. word-type) depending on
probabilities specified below, see Fig.~\ref{fig.model} for a sketch of the model. 
The total number of word-types is given by $N=N_c+N_{\bar{c}}$, where ($N_{\bar{c}}$) $N_c$ is the number of (non)core-words. 
The new word-token can either be a new word-type ($N \mapsto N+1$) with a probability $p_{\mathrm{new}}$ or an already existing word-type ($N \mapsto N$) with probability $1-p_{\mathrm{new}}$. 
In the latter case, a (previously used) word-type is attributed to the word-token at random with probability proportional to the number of times this word-type has occurred before. 
In the former case, the new word-type can either originate from a finite set of $N_c^{\mathrm{max}}$ core-words ($N_c\mapsto N_c+1$) with
probability $p_c$ or come from a potentially infinite set of noncore-words ($N_{\bar{c}}\mapsto N_{\bar{c}}+1$). In our simplest model we
consider $p_c$ to be a constant, i.e. $p_c^0\lessapprox 1$, which becomes zero only if all core-words were drawn ($N_c=N_c^{\mathrm{max}}$):
\begin{equation}\label{eq.map.pc}
p_c \left( N_c \right) = \left\{\begin{array}{ll}
p_c^0  &  \text{ if } N_c < N_c^{\mathrm{max}}, \\
0 &  \text{ if } N_c = N_c^{\mathrm{max}}. \\
\end{array}
\right.
\end{equation}
The final element of our model, which establishes the distinguishing aspect of core-words, is the dependence of $p_{\mathrm{new}}$ on $N$.
We choose $p_{\text{new}}$ (and $p_c$) to depend on $N$ and not on $M$ because an increase in $N$ necessarily reflects that fewer
  undiscovered words exist while an increase in $M$ is strongly affected by repetitions of frequently used words.
By definition, we think of core-words as necessary in the creation of any text and, therefore, the usage of a new core-word in a particular
text should be expected and thus not affect the probability of using a new (non-core) word-type in the future, i.e., $p_{\mathrm{new}}=p_{\mathrm{new}}(N_{\bar{c}})$.
On the other hand, if a noncore-word is used for the first time ($N_{\overline{c}} \mapsto N_{\overline{c}} + 1$) the combination of this
word with the previously used (core and non-core) words lead to a combinatorial increase in possibilities of expression of new ideas with the already used vocabulary and
thus to a decrease in the 
marginal need for additional new words~\cite{Petersen2012:1}. In our model, this argument suggests that $p_{\mathrm{new}}$
should decrease with $N_{\bar{c}}$. Taking these factors into account, we propose as an update rule for $p_{\text{new}}$ after each occurrence of a
new non-core word as
%
\begin{equation}\label{eq.map.pnew}
 p_{\mathrm{new}} \mapsto p_{\mathrm{new}} \left(1- \frac{\alpha}{ N_{\overline{c}} + s }\right),
\end{equation}
with the decay rate $\alpha > 0$ and the constant $s\gg 1$ which is introduced simply in order to damp the reduction of $p_{\mathrm{new}}$ for small $N_{\overline{c}}$ (for simplicity, we use $s=N^{\mathrm{max}}_c$).
The main justification for the exact functional form in Eq.~(\ref{eq.map.pnew}) is that it allows us to recover the empirical
  observations reported in Sec.~\ref{sec.emp}, as shown below. An alternative {\it a posteriori} justification will 
  be given at the end of this section and shows that Eq.~(\ref{eq.map.pnew}) can be interpreted as a direct consequence of an unlimited
  non-core vocabulary.

We now show how this model recovers Eqs.~(\ref{eq.modeldp}) and~(\ref{eq.modelH}).
We require that $1 - p_c^0 \ll 1$, which simply means that it is much more likely to draw core-words than noncore-words initially. In this case we can obtain approximately exact solutions for $N(M)$ in the two limiting cases considered in Eq.~(\ref{eq.modelH}). 
When $N \ll N_c^{\mathrm{max}}$, which implies $N_c,N_{\bar{c}} \ll N_c^{\mathrm{max}}$,  it follows from Eqs.~(\ref{eq.map.pc}) and (\ref{eq.map.pnew}) that $p_{\mathrm{new}}\approx \mathrm{const.}$ and therefore we trivially obtain that $N\sim M^{1}$. 
This case resembles the very beginning of the vocabulary growth, when most new word-types belong to the set of core-words. 
In the case $N \gg N_c^{\mathrm{max}}$, $p_c=0$ and $N\approx N_{\overline{c}}$ so that Eq.~(\ref{eq.map.pnew}) becomes in the continuum limit:
\begin{equation}\label{eq.alpha}
 \frac{\text{d}}{\text{d}N}p_{\text{new}}\left( N \right) = -\alpha \frac{p_{\text{new}}\left(N \right) }{N},
\end{equation}
from which it follows that $p_{\text{new}} \sim N^{-\alpha}$.

We now obtain the expected growth curve $N(M)$.
Notice that our model can be considered a biased random walk in $N$, which, as an approximation, can be mapped onto a binomial random walk by the coordinate transformation $N(M)$ such that $p_{\text{new}} \left( N \right) = p_{\text{new}} \left( N(M) \right)$. 
The resulting Poisson-Binomial process~\cite{Feller1968} can be treated analytically, e.g., the transformation $N(M)$ is then given by the average of the vocabulary growth:
\begin{equation}\label{eq.NM}
\begin{alignedat}{3}
 N(M) &= \int_0^M \text{d}M' p_{\text{new}}\left(M' \right) \\
      &= \int_{N(0)}^{N(M)} \text{d}N' \left| \frac{\text{d}M'}{\text{d}N'}\right| p_{\text{new}}\left(N' \right).
\end{alignedat}
\end{equation}Using $p_{\text{new}} \sim N^{-\alpha}$, this equation holds (self-consistently) by assuming a sub-linear growth for the vocabulary $N\sim M^{\lambda}$, where the relation $\lambda = \left( 1+ \alpha \right)^{-1}$ is established (for details see SM-Sec.~IV~\cite{SM}).
In accordance with Eq.~(\ref{eq.modelH}), we identify the following relation between the parameters: $N_c^{\mathrm{max}}=b$ and $\alpha=\gamma-1$.
The fitting parameters of Eq.~(\ref{eq.modeldp}) can thus be interpreted as: 
$b$ is the size of the core vocabulary and $\gamma$ controls the sensitivity of the probability of using a new word to the number of already used words in Eq.~(\ref{eq.alpha}).
 
Since the probability of usage for already used word-types is assumed to be proportional to the number of times it occurred before,
we guarantee that Eq.~(\ref{eq.modelH}) implies~(\ref{eq.modeldp})~\cite{Zanette2005:0}, meaning that the double scaling in the Zipf plot is also recovered from our generative model.
While the previous arguments show that the correct scalings are obtained by our model, in order to obtain an agreement with the data it is essential to: (i) use the normalization constant $C$ in order to determine the initial probability of finding a new word in Eq.~(\ref{eq.map.pnew}); (ii) re-scale the distribution using the threshold $n$ as $M/n$; and (iii) account for the disproportionally large weight of the first word-types (in the Zipf plot). 
Taking these points into account, direct simulations of the model in Fig.~\ref{fig.model} with the traditional parameters $b=b^*$ and $\gamma=\gamma^*$ lead to Zipf's and Heaps' curves, which resemble the original fits. See SM-Sec.~V for all details~\cite{SM}.

It is worth comparing the generative model with the model of random usage of words with fixed frequency, the ZE model discussed in the
previous section. 
While the ZE model allowed us to obtain Heaps' curves from Zipf's distributions (and vice-versa), in the generative model we
simultaneously obtain the double scaling regime in both cases.
It is important to stress that individual texts or single databases should not be considered as the output of single realizations of our
generative model. Instead, we consider that not only texts but also all databases 
have a negligible size when compared to the language as a whole and therefore should be thought of as a small 
subsample  ($M_{\mathrm{database}} \ll M$) of the output of our generative model, retrieved after it achieved its stationary state ($M \rightarrow
\infty$). In this case, changes in word frequencies become negligible (in the scale of $M$) during the creation of the
database (in the scale of $M_{\mathrm{database}}$).
Therefore, the vocabulary growth of the created database is well approximated by the ZE model with $F_{\text{dp}}(r)$.

Finally, we take profit of our previous calculations and provide an {\em a posteriori} justification of the key assumption of our model,
  Eq.~(\ref{eq.map.pnew}).
 Our starting point is the observation -- see Fig.~\ref{fig.Heaps}(d) -- that vocabulary is for all practical purposes infinite. We therefore 
 postulate that 
\begin{equation}\label{eq.postulate}
 N(M) \overset{M \rightarrow \infty}{\longrightarrow} \infty,
\end{equation}
and by following (in reverse order) the previous calculations  we naturally arrive at Eq.~(\ref{eq.map.pnew}).
 From the first line of Eq.~(\ref{eq.NM})
we see that in order to fulfill our postulate~(\ref{eq.postulate}),  $p_{\text{new}}$ has to decay at least as slow as $p_{\text{new}}\left(M \right) 
\sim M^{-\delta}$ with $\delta \le 1$ for $M\rightarrow \infty$. In a minimal model it is reasonable to assume such a power-law decay, in
which case the first 
line of Eq.~(\ref{eq.NM}) implies that $N(M) \sim M^{\lambda}$ with $\lambda = 1-\delta$.  Making a transformation of variables from $M$ to
$N$ we obtain
\begin{equation}
 p_{\text{new}}\left( N \right) = p_{\text{new}}\left( M(N) \right) \sim N^{-1+\frac{1}{\lambda}}=N^{-\alpha}.
\end{equation}
In turn this is equivalent to Eq.~(\ref{eq.alpha}), from which we recover Eq.~(\ref{eq.map.pnew}) as a discretized version.
Thus we see that Eq.~(\ref{eq.map.pnew}) is a minimal assumption for an unbounded vocabulary.

\section{Historical changes}\label{sec.hist}
\begin{figure*}[bt]
\centering
\includegraphics[width=2\columnwidth]{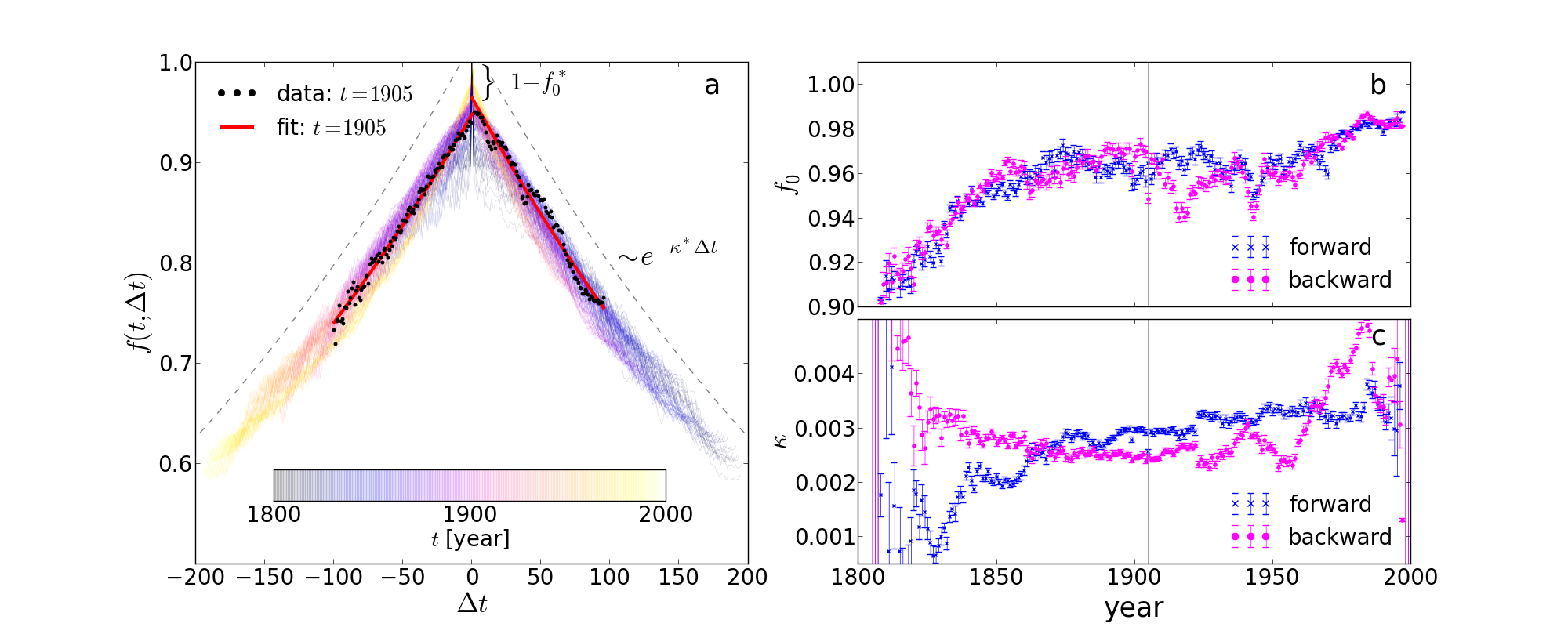}
\caption{Historical change in the composition of core-words in the English vocabulary. a) fraction $f(t,\Delta t)$ of core-words in  $y(t)$ that remain in the set of core-words in $y(t+\Delta t)$ for $t\in [1805,2000]$ (pale colors) and in particular for $t=1905$ (black dots) with the corresponding exponential fit (red line).
b+c) Parameters $f_0$ and $\kappa$ in the exponential decay Eq.~(\ref{eq.hist.decay}) of the curves in a) obtained through least-square fits. Forward (backward) decay refers to $\Delta t>0$ ($\Delta t<0$).  }
\label{fig.Cores}
\end{figure*}  
The model described so far has been shown to give a good account for all databases and all years with the same fixed two parameters $N_c^{\text{max}}=b^* =7,873$ and $\alpha=\gamma^* - 1 = 0.77$ in the case of English.
A natural question is, therefore, what actually changes in historical time scales? 
Considering two different databases (say two different years), our model does not consider any differences in the actual composition of the core-vocabulary.
Even if the value of $N_c^{\text{max}}$ remains constant this does not mean that the {\em same} words are observed for all years. 
From the point of view of our model, the main change a word can experience is to enter or to leave the group of core-words. 
For instance, comparing the decades $1891-1900$ and $1991-2000$, the most frequent words which left the core-vocabulary were {\it
    majesty}, {\it doubtless}, {\it furnished}, {\it monsieur}, {\it napoleon}, and {\it hitherto}, while the ones which entered were {\it
    cultural}, {\it context}, {\it technology}, {\it programs}, {\it environmental}, and {\it computer}~\cite{footnote:ex}

In order to quantify this effect, we investigate the replacement of words from the core-vocabulary in the yearly databases $y(t)$ in the time $t \in [1805,2000]$ in Fig.~\ref{fig.Cores}.
We calculate the fraction $f(t,\Delta t)$ of core-words (i.e. with rank $r<b^*=7873$, fixed for all $t$) from $y(t)$ that remain in the set of core-words in $y(t+\Delta t)$.
Figure~\ref{fig.Cores}(a) shows that all curves can be qualitatively described by an exponential decay 
\begin{equation}\label{eq.hist.decay}
 f(t,\Delta t) = f_0 e^{-\kappa \left| \Delta t \right|},
\end{equation}
independent of whether forward ($\Delta t>0$) or backward time ($\Delta t<0$) was considered.  
This is further supported in Fig.~\ref{fig.Cores}(b-c), where the parameters $f_0$ and $\kappa$ obtained numerically from a least-square fit~\cite{Press2007} of Eq.~(\ref{eq.hist.decay}) for all curves $f(t,\Delta t)$ with $t \in [1805,2000]$ are presented.
In order to avoid biases due to different number of points in the fit, for each $t$ we performed a fit with the same number of points $\min\{2000-t,t-1805\}$ forwards and backwards in time. 
On closer inspection, two features connected to the interpretation of the parameters $f_0$ and $\kappa$ deserve a more careful discussion.
The parameter $f_0<1$ represents the discontinuous change of core-words in two subsequent years.
It strongly depends on the different selection of books in the construction of the respective databases and can be attributed to the finite size of the database, which leads to a wrong estimation of the ``true'' core-words.
Consistently with this interpretation, Fig.~\ref{fig.Cores}(b) shows that $f_0$ grows over time, due to the fact that database size increases leading to a better sampling of words. 
Nevertheless, a value of $f_0 \approx 0.98$ indicates that this is still far from being negligible (e.g., for $N_c^{\text{max}} = 7,873$ this means that around $150$ words of the set of core-words will be different due to finite sampling).
In contrast, the decay rate $\kappa$ describes the continuous replacement of core-words over time with a rate of $\kappa N_c^{\text{max}} \approx 30$ words per year. 
The most intriguing observation in Fig.~\ref{fig.Cores}(c) is that this change experiences an acceleration over time as $\kappa$ grows by more than $50\%$ from $1805$ to $2000$.

Finally, it is worth discussing the implications of these findings on our generative model. The characteristic time scale of the core-vocabulary replacement ($\approx 1/\kappa$) is on the order of centuries. This means that on the scale of a few decades our generative model holds with the asumption of a constant core vocabulary. On longer time scales our model has to be refined in order to include: i) a probability of replacement of the words belonging to the core-vocabulary; and ii) a finite memory or a distinction between core- and noncore-words in the preferential attachment part of our model.

\section{Discussion}

In summary, we have shown that the rank frequency distribution and the vocabulary growth of languages can be best described by simple two-scaling
functions. The only two free parameters of the functions are related to each other and remain almost unchanged over centuries as well as databases and depend only on the considered
language. 
We have also shown that these empirical findings can be interpreted as the result of a finite number of words belonging to a core vocabulary, which have different properties from the remaining virtually unlimited number of words, as summarized in Tab.~\ref{tab.core}. 
This conclusion was achieved based on a simple generative stochastic model for the vocabulary growth. 
Finally, we found that in English the composition of the core-vocabulary experiences an exponential decay with a rate of $30$ words per year, which is, remarkably, steadily accelerating in the past decades. 
\begin{table}[ht]
\begin{tabular}{|c||c|c|}
\hline
    & Core  Words & Non-core-words \\
\hline
\hline
Number & finite: $N_c^{\mathrm{max}} \in [10^3,10^4]$ & infinite: $N_{\bar{c}}\rightarrow \infty$ \\
\hline
Frequency & larger ($r>b^*$) & smaller ($r<b^*$) \\
\hline
Effect on $p_{\mathrm{new}}$ & none & reduction \\
\hline
\end{tabular}
\caption{Properties of core ($c$) and noncore ($\bar{c}$) words in our model.}
\label{tab.core}
\end{table}

It is worth comparing these findings in view to previous results. 
As far as we are aware, our analysis provides the first rigorous statistical confirmation of similar previous proposals~\cite{Naranan1998:0, Cancho2001:1, Petersen2012:1} of the double-scaling generalizations of Zipf's law -- Eq.~(\ref{eq.modeldp}). 
The consequence of this to vocabulary growth and Heap's law (see also~\cite{Petersen2012:1}), which we drew based on a Poisson usage of words~\cite{Eliazar2011:1}, is that the rate of introduction of new words decays but never vanishes with increasing database size. 
This is in contrast to recent claims which reported a convergence to a maximum vocabulary size~\cite{Bernhardsson2009:0}. 
We note that this previous analysis was based on single books and therefore the database sizes were close to our transition point $N_c^{\mathrm{max}}$, which we believe could have been misinterpreted as a systematic decay. 
A generalization of a Yule's type process to obtain double-scaling degree distribution in a network of words was introduced in Ref.~\cite{Dorogovtsev2001:0}. 
Two crucial differences to our model are that it yields fixed exponents and cannot be understood as a generative model of texts (word by word). 
Interestingly, in Ref.~\cite{Levary2012:0} an analysis of the network constructed from the thesaurus also showed the existence of a set of core-words almost of the same size as ours.

Our simple model and expression for the vocabulary growth as a function of database size has important practical consequences. 
Simply knowing the database size (in number of words, $M$, or potentially in bits), and using the language dependent parameters ($C,N_c^{\mathrm{max}}=b^*,\alpha=\gamma^*-1$) reported above, from Eq.~(\ref{eq.modelH}) one can immediately estimate  the expected number of different words, $N$, appearing more than $n$ times.
This is crucial for search engines and data mining programs because it allows for an estimation of the memory to be allocated prior to the scanning of an unknown database, e.g., in the construction of the inverted index~\cite{Baeza-Yates2000:0, Williams2005:0, Croft2009}.
Even the fluctuations around this expectation can be easily computed through our generative model or through the Poisson assumption of word-usage. 
Of course, this strong assumption ignores correlations and typically underestimates the expected fluctuations, so that our model should be considered as the simplest null model. 
The existence of a transition between two scalings (which is under the reach of even single large books) shows that simple estimations based only on the
traditional Zipf's law have to be generalized. 
For instance, a commonly used index of vocabulary richness of a text is Herdan's coefficient given by the ratio $\log N/\log M$~\cite{Baayen2001}.
In view of our results, the coefficient is highly dependent on which of the two scaling regimes is reached with the given size of the text. 

We now compare our observations of change on historical time scales to other historical changes in language usage. 
For the whole vocabulary, we obtain that the vocaulary size is mainly driven by the available database size. 
This is in contrast to previous conclusions based on the same google-ngram database which detected a growth of vocabulary in time~\cite{Michel2011:0}. 
Here it is important to note that this previous analysis included a substantially different filtering of the listed 1-grams to achieve valid words in the vocabulary, including a frequency criterion and manual classification. 
Still, our results show that also in this case a more careful analysis of the role of the database size is needed. 
For the core vocabulary, we observe a fairly constant number of constituents over centuries. 
The number of words common to core-vocabularies of different databases was found to decay exponentially with the time between publication of the databases, e.g., for English the decay rate is approximately $30$ words per year and the half-life of the core vocabulary is $\approx 200$ years. 
It is worth to compare these numbers with recent studies which reported half-lifes for: i) the regularization of verbs ($750$ to $10\,000$ years)~\cite{Lieberman2007:0}, and ii) a fundamanetal vocabulary of 200 words ($300$ to $38\,000$ years)~\cite{Pagel2007:0}.
Perhaps our most intriguing finding is the approximately linear increase of the rate in time, which eventually confirms the overall acceleration of language change and society in general, as propagated in Ref.~\cite{Michel2011:0}.

Our results can be extended in many directions and open new possibilities of studies of vocabulary change. 
Directly related to our observations and model, it remains to be explained the specific value of the parameter $\gamma^*\approx 1.77$, which is intriguingly similar across different languages. 
Another important point is to assess the limitations of our estimations due to the role of correlations inside real texts and databases, and how this could be introduced into our model.
Furthermore, it remains to be shown whether the transition between two scalings due to the existence of a core vocabulary can be related to the phenomenon of phase transitions in ranking stability of complex systems recently reported in Ref.~\cite{Blumm2012:0}.
Finally, we believe that our model provides the correct null model for normalizations due to database sizes and that therefore future investigations of historical effects on the vocabulary should take this into account. 

\acknowledgments
We are indebted to J. Miotto, and R. Guimer\`a for valuable discussion about the data analysis. We thank F. Ghanbarnejad and J. Leit\~ao for
the careful reading of the manuscript.

%
\newpage
\clearpage
\begin{widetext}

\begin{center}
{\large {\bf Supplemental Material for: \\ Stochastic model for the vocabulary growth of natural languages \vspace{15pt}}} \\ 
Martin Gerlach$^{1}$ and Eduardo G. Altmann$^{1}$ \\
$^{1}$ \textit{Max Planck Institute for the Physics of Complex Systems, 01187 Dresden, Germany}
\end{center}

\renewcommand{\theequation}{S\arabic{equation}}
\renewcommand{\thefigure}{S\arabic{figure}}
\renewcommand{\thetable}{S\arabic{table}}
 
\setcounter{equation}{0}  
\setcounter{figure}{0}
\setcounter{table}{0}
\setcounter{section}{0}

\maketitle

\section{Data}\label{sec.data}
The data obtained from the google-ngram database~\cite{SMichel2011:0} is filtered in two steps. 
First, we decapitalize each word (e.g. 'the' and 'The' are counted as the same word) and further restrict ourselves to words consisting uniquely of letters present in the alphabet of the corresponding language and the symbol `` ' '' (apostrophe).
This is meant as a conservative approach in order to minimize the influence of foreign words, numbers (e.g. prices), or scanning problems which are present in the raw data.
In the second step, when constructing yearly data $y(t)$, i.e., words present in books published in year $t$, we include only those words in the database $y(t)$, which appear more than $40$ times in that particular year. 
In the same way, for the cumulative data $Y(t)$ we include only those words, which appeared more than $40$ times until time $t$.
In this way we avoid a possible bias due to the filtering applied in the construction of the raw data (words had to appear more than $40$ times in all times in order to be included in the database~\cite{SMichel2011:0}).
As an example of possible bias, in case we had not applied this filter, take two words (called '$1$' and '$2$') with $N_1(t) = N_2(t)=21$ occurrences in year $t$. 
If now $\forall t'\neq t:N_1(t')=0$ and $\exists t''\neq t:N_2(t'')>20$, word '$2$' would be present in the raw data whereas word '$1$' would be not.
As a result we would only include word '$2$' in the yearly database $y(t)$.
With our additional filter neither word '$1$' nor word '$2$' appears in the yearly database $y(t)$.

In Fig.~\ref{fig.Size} we show the resulting database size for the yearly data $y(t)$ and the cumulative data $Y(t)=\sum_{t'=to}^{t} y(t')$ in terms of word-tokens and word-types for English, French, Spanish, German, and Russian.
In this context word-type refers to the number of distinct words, whereas word-token refers to the total number of words.

For the yearly database $y(t)$ we use data in the period $t \in [1805,2000]$, because as already indicated in~\cite{SMichel2011:0}, the database composition may have changed in a noncontinuous way at $t \approx 1800$.
This claim is supported in Fig.~\ref{fig.Rank}, where we calculate Kendall's rank correlation coefficient $\tau[y(t),y(t')]$ between the common types of the database $y(t)$ and $y(t')$ for $1500\leq t\leq t' \leq2000$ as
\begin{equation}
 \tau[y(t),y(t')] = \frac{n_c - n_d}{\frac{1}{2} n(n-1)},
\label{eq.Kend.tau}
\end{equation}
where $n$ is the total number of common elements, $n_c$ the number of concordant, and $n_d$ the number of disconcordant pairs between the two databases with respect to the ranking of frequencies.
Clearly, at $t=1800$ a noncontinuous change in $\tau$ can be identified, from which we conclude that database composition is dramatically different in the years before and after $t=1800$. 
In order not to be affected by this change the yearly data $y(t)$ is only considered in the period $t \in [1805,2000]$.
However, in order to take advantage of the full size of the database, the cumulative data $Y(t)$ is constructed taking into account all the years prior to $t=1805$.

\section{Maximum Likelihood Estimation}\label{sec.ml}
\subsection{Theory}\label{sec.ml.th}
In this section we give account of the distributions proposed for fitting the rank-frequency distribution and present the details of the Maximum likelihood estimation procedure. 
The procedures are standard~\cite{SPress2007}, but here we fit directly the rank frequency distribution originally proposed by Zipf~\cite{SZipf1936} instead of the word frequency distribution considered in Ref.~\cite{SNewman2005:0}.

\begin{table}[ht] 
\begin{tabular}{|c||l|c|c|}
\hline
 i & distribution  & $F(r;\Omega)$ & set of parameters $\Omega$ \\
\hline
\hline
1 & Power-Law & $C r^{-\gamma}$ & $\gamma$ \\
\hline
2 & Shifted Power-Law & $C (r+b)^{-\gamma}$ & $\gamma$, $b$ \\
\hline
3 & Power-Law with Exponential cutoff (tail) & $C \exp \left({-br}\right)r^{-\gamma}$ & $\gamma$, $b$ \\
\hline
4 & Power-Law with Exponential cutoff (beginning) & $C \exp \left({-b/r}\right)r^{-\gamma}$ & $\gamma$, $b$ \\
\hline
5 & Log-normal & $C r^{-1} \exp \left(-\frac{1}{2}\left(\ln r - \mu \right)^2 /\sigma^2 \right) $ & $\mu$, $\sigma$ \\
\hline
6 & Weibull & $Cr^{\gamma -1} \exp \left(-br^{-\gamma} \right)$ & $\gamma$, $b$ \\
\hline
7 & Double Power-Law & $C\begin{cases}
r^{-1}, & r\leq b\\
b^{\gamma-1}r^{-\gamma} & r>b,
\end{cases}$ & $\gamma$, $b$ \\
\hline
\end{tabular}
\caption{Proposed models to fit rank-frequency distributions.}
\label{tab.distmodels}
\end{table}
In Tab.~\ref{tab.distmodels} the proposed descriptive models used to fit the rank-frequency distribution are presented.
The notation $F(r;\Omega)$ means that the distribution $F$ depends on the rank $r$, and $\Omega$ is the set of parameters.
The normalization constant $C=C(\Omega)$ is a function of the respective parameters and fixed by $\sum_{r=1}^{\infty} F(r;\Omega)=1$.
In practice, this is calculated with the Euler-Maclaurin formula available in the package mpmath~\cite{Smpmath2010}.

The parameters of each distribution are estimated numerically by minimizing the negative of the log-Likelihood 
\begin{equation}
 \Omega^* = \arg \underset{\Omega}{\min} \mathcal{L}'\left( \Omega \right),
\end{equation}
where
\begin{equation}
 \mathcal{L}'\left( \Omega \right) = -\ln \mathcal{L} = -\sum_{i=1}^M \ln F \left( r\left(i\right);\mathrm{\Omega}\right).
\end{equation}
In this expression $M$ is the number of tokens, which implies that the sum goes over each observed token $i$ and its corresponding rank $r(i)$.
In practice, the minimization is obtained with a Nelder-Mead simplex algorithm (available in the Scipy library~\cite{Sscipy}).

The quality of the fit was evaluated quantitatively by means of a $p$-value obtained from a $\chi^2$-statistics~\cite{SD'Agostino1986}:
\begin{equation}
 \chi^2 = \sum_{j=1}^Q =  \frac{\left(N_j - n_j\right)^2}{n_j}.
\end{equation}
Here the domain is partitioned into $Q$ cells, such that the expected number of observations per cell $n_j \geq 5$~\cite{STaylor1997}, with $N_j$ being the actual observed number of observations in cell $j$.
A recently proposed alternative strategy~\cite{SClauset2009:0} involving the comparison of the Kolmogorov-Smirnow statistics of the actual empirical data with randomly generated data is computationally not feasible in this case, because it would require us to draw $\approx 10^{15}$ random numbers ($p$-value precision $0.01$)  due to the size of the database of $>10^{11}$ tokens.

In the last step we determine which of the proposed models $i=1...R$, where $R$ is the number different models considered, is most likely to describe the data.
In order to account for the different number of fitted parameters we calculate the Akaike information criterion (AIC)~\cite{SAkaike1974:0} for each model $i$
\begin{equation}
 AIC = 2 \mathcal{L}' \left( \Omega^* \right) + 2K,
\end{equation}
where $K$ is the number of parameters estimated in the model.
The model which gives the minimum value $AIC_{\mathrm{min}}=\underset{i}{\min} \{ AIC_i\}$ is most likely to describe the given data.
From this we can calculate the relative likelihood $l_i$~\cite{SBurnham2002}
\begin{equation}
 l_i = \exp \left( -\left(AIC_i-AIC_{\mathrm{min}}\right)/2 \right),
\end{equation}
which states how likely model $i$ is to describe the data in comparison with the best model.
This implies that the probability $w_i$ that model $i$ (out of the $R$ models considered) describes the data is given by~\cite{SBurnham2002}
\begin{equation}\label{eq.wi}
 w_i=P(\mathrm{model}\,i|\mathrm{data}) = l_i/ \sum_{j=1}^{R} l_j.
\end{equation}

\subsection{Results}\label{sec.ml.res}
\subsubsection*{Google-ngram}
In this section we give a detailed overview of the results obtained from fitting the models in Tab.~\ref{tab.distmodels} to the rank-frequency distributions for all languages considered, i.e., English, French, Spanish, German, and Russian.
In Fig.~\ref{fig.AIC.eng} - \ref{fig.AIC.rus}(a+b) we plot the $AIC$ from the models in Tab.~\ref{tab.distmodels} applied to yearly $y(t)$ and cumulative data $Y(t)$ of the respective language. In Fig.~\ref{fig.AIC.eng} - \ref{fig.AIC.rus}(c) we show explicitly the rank-frequency distribution of the data $Y(2000)$ and the corresponding fits of the three models that yield the best description: the double power-law ($i=7$), the power-law with an exponential cutoff in the tail ($i=3$), and the log-normal ($i=5$).  

For English, $i=7$ yields the best description of the yearly data for $t\gtrsim1950$ and for the cumulative data for $t\gtrsim 1810$.
As the databases $y(1950)$ and $Y(1810)$ can be considered independent datasets and by comparing with Fig.~\ref{fig.Size}(a) we conclude that the size of the database needs to exceed a certain threshold ($\approx 10^9$ tokens) in order to discriminate competing models like the $i=3$ in the tail. 
This is further corroborated by looking at the inset in Fig.~\ref{fig.AIC.eng}(c), where it can be seen that $i=7$ outperforms $i=3,5$ especially in the description of the tail of the distribution.

For the other languages except English the $AIC$ of the yearly data $y(t)$ favours $i=3$.
This comes with no surprise since their size is limited to $<10^9$ tokens for all $t\in [1805,2000]$, as can be seen in Fig.~\ref{fig.Size}(a).
In contrast, the cumulative data $Y(t)$ shows different results.
For French and Spanish the $AIC$ favors $i=7$ as the size of the database grows, especially for the largest dataset $Y(2000)$. 
Again, this becomes clear when looking at the deviations of the fits to the real data in the inset of Fig.~\ref{fig.AIC.fre}(c),~\ref{fig.AIC.spa}(c), which seem to diverge for $i=3,5$ in the tail of the distribution.
For German and Russian the $AIC$ identifies $i=7$ only as the second best fit for the cumulative data $Y(t)$.
This is most probably due to the fact that the size of the database for those languages is still not large enough in order to discriminate a second power-law regime clearly.
Additionally, for these languages the critical rank $b^*$, where a transition between the two power-laws occurs, is shifted towards higher values, possibly due to the different degree of inflection (see main text).
This in turn implies that the fraction of tokens belonging to the power-law in the tail is much smaller than in English, which means that a larger database is needed in order to discriminate $i=3,5$. 
This claim is further supported by the insets of Fig.~\ref{fig.AIC.ger}(c),~\ref{fig.AIC.rus}(c), where we show that especially in the tail of the distribution $i=7$ deviates less from the data than the competing models.

Whereas English, French, and Spanish give approximately the same values for the largest database $Y(2000)$, German and Russian show larger values for $b$ and a different power-law exponent in the tail (see main text). 
The latter might point towards more subtle differences between the languages besides inflection.

\subsubsection*{Wikipedia}
In this section we want to show that our findings related to the double power-law fit are indeed of general validity and do not originate from peculiarities  of the google-ngram database, e.g. scanning problems. For this we choose a complete snapshot of the English Wikipedia~\cite{SWikidump2012}, because i) it contains a large amount of text, ii) the text does not need to be scanned, and iii) the publishing process is inherently different from that of books.

We filter the Wikipedia database in three steps.
First, using the WikiExtractor developed by the University of Pisa Multimedia Lab~\cite{SWikiExtractor}, we store only the plain text, neglecting any additional information or annotation such as images, tables, tags, references, or lists.
In a second step we remove all punctuation characters (e.g. ``,'', ``;'', or ``\{'') and cut the text into words at the whitespace characters in a similar manner as described in the construction of the google-ngram database~\cite{SMichel2011:0}.
The final step consists of decapitalizing each word and restricting ourselves only to words consisting uniquely of the letters $a-z$, a filter we also applied to the google-ngram database (see SM-Sec.~\ref{sec.data}).
The resulting sequence of words consists of $N \approx 3.7\,\, 10^6$ types and $M\approx1.3\,\, 10^9$ tokens.
\begin{table}[b] 
\begin{tabular}{|c||l|c|}
\hline
 i & distribution  & $AIC/M$  \\
\hline
\hline
1 & Power-Law & $15.972$  \\
\hline
2 & Shifted Power-Law & $15.782$  \\
\hline
3 & Power-Law with Exponential cutoff (tail) & $15.662$  \\
\hline
4 & Power-Law with Exponential cutoff (beginning) & $15.821$  \\
\hline
5 & Log-normal & $15.574$  \\
\hline
6 & Weibull & $17.740$  \\
\hline
7 & Double Power-Law & $15.525$  \\
\hline
\end{tabular}
\caption{Values $AIC/M$ from fitting the proposed models of Tab.~\ref{tab.distmodels} to the rank-frequency distributions of the English Wikipedia with $M = 1\,257\,349\,755$ tokens.}
\label{tab.wikifit}
\end{table}

Following the recipe in SM-Sec.~\ref{sec.ml.th}, we show that the results for fitting the models in Tab.~\ref{tab.distmodels} to the rank-frequency distribution of the Wikipedia database is consistent with the results from the google-ngram database. In Tab.~\ref{tab.wikifit} we show the values for the $AIC$ from which we can see that the double power-law is the best fit among the proposed models with a probability $1-p<10^{-15}$.
Additionally, in Fig.~\ref{fig.AIC.wikien} we plot the rank-frequency distribution of the Wikipedia data and the corresponding fits of the three models that yield the best description: the double power-law ($i=7$), the power-law with an exponential cutoff in the tail ($i=3$), and the log-normal ($i=5$). 
This corroborates our claim that the double power-law is the best fit for the rank-frequency distribution.
Furthermore, the estimated values for the parameters are $\gamma=1.68$ and $b = 7830$, which closely matches our observations from the google-ngram database ($\gamma^*=1.77$, $b^*=7873$).

\section{Zipfian Ensemble}\label{sec.ze}
\subsection{Theory}\label{sec.ze.th}
The Zipfian Ensemble (ZE)~\cite{SEliazar2011:1} is a simple approach to model the size of the vocabulary depending on the text length given the rank-frequency distribution $F(r)$, $r=1...N_{\mathrm{ZE}}^{\mathrm{max}}$, where $N_{\mathrm{ZE}}^{\mathrm{max}}\in [1,\infty)$ is the hypothetical (maximum) size of the vocabulary.
The occurrence of each word-type with rank $r$ is assumed to be governed independently by a Poisson process with an intensity equal to its frequency, e.g., $\lambda(r)=F(r)$, where time is measured in units of tokens $M$ (text length).
This means that the probability that this word-type occurs at least once in the interval $T_1 \in [0,M]$ is given by~\cite{SFeller1968}
\begin{equation}
 P\left( T_1 \leq M;r \right)=1-e^{-F(r)M}.
\label{eq.ZE1}
\end{equation}
From this we can calculate the vocabulary size $N(M)$ by summing over all word-types, which gives the expected (average over realizations) number of words (out of $N_{\mathrm{ZE}}^{\mathrm{max}}$ different words in total) that appeared at least once up until time $M$:
\begin{equation}
 N(M) = \sum_{r=1}^{N_{\mathrm{ZE}}^{\mathrm{max}}} \left[ 1 - e^{-F(r)M} \right].
\label{eq.ZE2}
\end{equation}
The variance of the ZE over the different realizations indicates the expected fluctuation around $N(M)$ in Eq.~(\ref{eq.ZE2}) and is given by \cite{SEliazar2011:1}:
\begin{equation}
 \mathbb{V}\left[ N(M) \right] = N(2M) - N(M).
\end{equation}
 Although similar, this framework differs from the usual 'bag-of-words' (or shuffled texts) in the sense that i) the expected time of occurrence of a word need not to be an integer and ii) two words can in principle occur at the same time due to the independence of the Poisson processes. This in turn limits the interpretation of the ZE as a model for the creation of a text token by token.
However, it allows for an analytic treatment and the continuous time approximation becomes better in the limit of large databases.

\subsection{ZE in the double power-law}\label{sec.ze.dp}
In this section we want to show that a double power-law in the rank frequency distribution (Eq. (1) main text) can lead to the double scaling in the vocabulary growth (Eq. (2) in main text) in the framework of the ZE.

First, we generalize the ZE to cases where words have to appear at least $n$ times before they are considered part of the vocabulary.
The introduction of a threshold $n$ means that instead of looking at the probability for the time until its first occurrence $T_1$, one considers $T_n$, the time it takes until the word occurs $n$ times and Eq.~(\ref{eq.ZE1}) becomes
\begin{equation}
 P\left( T_n \leq M;r \right)=1-\sum_{j=0}^{n-1}\frac{\left( F(r)M \right)^{j}}{j!}e^{-F(r)M}.
\label{eq.ZE1n}
\end{equation}
From this, Eq.~(\ref{eq.ZE2}) can be directly extended to
\begin{equation}
 N^{(n)}(M) = \sum_{r=1}^{N_{\mathrm{ZE}}^{\mathrm{max}}} \left[1 - \sum_{j=0}^{n-1}\frac{\left( F(r) M \right)^j}{j!}e^{-F(r)M}\right].
\label{eq.ZE2n}
\end{equation}

In the next step we consider the limit $n\gg 1$. 
As the stochastic variable $T_n$ is the sum of $n$ times the stochastic variable $T_1$, which is distributed according to Eq.~(\ref{eq.ZE1}), one can conclude that by means of the central limit theorem it follows that $P\left( T_n = M/n;r \right)$ will approach a Gaussian with vanishing variance, such that by rescaling $M\mapsto M/n$ Eq.~(\ref{eq.ZE1n}) asymptotically becomes 
 \begin{equation}
 \underset{n \rightarrow \infty}{\lim} P\left( T_n \leq M/n;r \right)=\Theta \left(M/n - \tau (r) \right),
\end{equation}
where $\tau(r) = 1/F(r)$ is the inverse of the frequency $F(r)$ of the particular word-type and $\Theta(x)$ is the Heaviside step function.
For the vocabulary growth this yields
 \begin{equation}
 \underset{n \rightarrow \infty}{\lim} N^{(n)}(M/n) = \sum_{r=1}^{N_{\mathrm{ZE}}^{\mathrm{max}}} \left[1 - \Theta \left(M/n - \tau (r) \right) \right] .
\end{equation}
Thus we obtain a direct relationship between the rank-frequency distribution and the vocabulary growth 
 \begin{equation}
 \underset{n \rightarrow \infty}{\lim} N^{(n)}(\tilde{M} = 1/F(r)) = r,
 \label{eq.ZEasym}
\end{equation}
where $\tilde{M}=M/n$.

In Fig.~\ref{fig.Heapskmin.ze} we show the $N^{(n)}(\tilde{M})$ curve obtained from the ZE for the double power-law (Eq.(1) main text) with parameters $\gamma^*=1.77$ and $b^*=7873$ for different thresholds $n$. 
One can see that the growth curves for $n>8$ are almost indistinguishable from the asymptotic solution Eq.~(\ref{eq.ZEasym}), which can be attributed to the fast convergence implied by the central limit theorem.

From these observation we conclude that Eq.~(\ref{eq.ZEasym}) is already a good approximation for $n\gg 1$, where in practice this can mean $n>10$.
As a result we obtain Eq. (2) from the main text. 
This means that the increase of the threshold $n$ leads to a reduction of the fluctuations of the growth curve of the vocabulary and can be explained as a result of a simple stochastic process. 
In Fig.~\ref{fig.Heapskmin.books} we show that this claim holds when applied to real texts of the size of single books, as well as for a collection of several million books, as in Fig.~\ref{fig.Heapskmin.data}.

\section{Selfconsistent Solution for Eq. (6)}
In this section we investigate the selfconsistent solution of Eq.~(6), main text, 
\begin{equation}\label{eq.Eq(6)}
 N(M)= \int_{N(0)}^{N(M)} \text{d}N' \left| \frac{\text{d}M'}{\text{d}N'}\right| p_{\text{new}}\left(N' \right).
\end{equation}
Assuming $p_{\mathrm{new}}=c_2 N^{-\alpha}$, we show that $N(M)=c_1 M^{\lambda}$ yields a true statement of Eq.~(\ref{eq.Eq(6)}) only when $\lambda = (1+\alpha)^{-1}$.
For the selfconsistent solution $N(M)=c_1 M^{\lambda}$ the Jacobian in Eq.~(\ref{eq.Eq(6)}) gives
\begin{equation}
 \frac{\mathrm{d}M}{\mathrm{d}N} = \frac{1}{\lambda c_1} \left( \frac{N}{c_1} \right)^{\frac{1}{\lambda} -1}.
\end{equation}
Noting that $N(0) = 0$, this gives for Eq.~(\ref{eq.Eq(6)}):
\begin{equation}
\begin{alignedat}{3}
 N(M) &=  \int_0^{N(M)} \text{d}N' \frac{1}{\lambda} c_2 c_1^{-\frac{1}{\lambda}} N'^{\frac{1}{\lambda}-\alpha-1} \\
      &=  \frac{1}{\lambda} c_2 c_1^{-\frac{1}{\lambda}} \frac{1}{\frac{1}{\lambda}-\alpha} N(M)^{\frac{1}{\lambda}-\alpha}.
\end{alignedat}
\end{equation}
We find that this equation holds only if $\lambda = (1+\alpha)^{-1}$ and $c_2 = \lambda c_1^{\frac{1}{\lambda}}$.

\section{Numerical simulation of the stochastic model}\label{sec.rwsim}
In this section we show the results of the direct numerical simulation of the model proposed in Sec.~III, main text. 

\subsection{Parameters and initialization}\label{sec.rwsim.par}
In order to simulate the model, apart from fixing a number of parameters ($N^{\textrm{max}}_c,\alpha,p_c^0$), we need to prescribe how the model is
initialized, e.g., what is the initial probability of using a new word $p_{\mathrm{new}}^0$ and how many word types exist at the first
iteration of the model. Concerning the parameters, the initial probability of choosing a core-word is set to $p_c^0=0.99$, such that
$1-p_c^0 \ll 1$ (see main text) and the two other parameters are fixed by the fitting parameters ($N_c^{\mathrm{max}}=b^*=7873$,
$\alpha=\gamma^*-1=0.77$ in English, see main text). Concerning the initialization of the model, an important point that needs to be taken into account is
that we are interested in retrieving the Heaps' plot obtained after re-scaling the number of tokens $M$ by the the
threshold $n$ as $\tilde{M}=M/n$, see Fig.~\ref{fig.Heapskmin.data} (for simplicity and computational efficiency in our simulations we
choose $n=1$). This implies that the first word type of our model should on average appear not at the first token but instead approximately at $\tilde{M}
\approx 1/F(1)$ (where $F(1)$ is the frequency of the most frequent word). In view of this requirement, we set
$p_{\mathrm{new}}^0=C=F_{dp}(1)=0.0922$ (for English, see main text) and we start with an empty list of word types (the tokens used before the appearance of the first
word type are counted but not attributed to any word type). The simulations were done with a maximum number of $M=10^9$ steps in units of
tokens, a restriction imposed by the computational effort required. The reported results were obtained as the average of $100$ realizations
of the model.

\subsection{Heaps' plot}\label{sec.rwsim.h}

In order to be able to compare the results with the google-ngram data, where a natural threshold of $n=41$ is imposed (see
SM-Sec.~\ref{sec.data}), we incorporate the threshold $n$ by using the rescaled coordinate $\tilde{M} = M/n$, as motivated and discussed in
SM-Sec.~\ref{sec.ze.dp}. In Fig.~\ref{fig.SimRW_h} we show the expected vocabulary growth $N(\tilde{M})$.
We can clearly see that the two scaling regimes of Eq.~(2), main text, are recovered from our model. Deviations from the data are within
$50\%$ over as much as $7$ orders of magnitude. The poorer agreement for large $\tilde{M}$ can be attributed to a slight overestimation by
our model of the point of transition between the two scaling regimes. This could be addressed by
modifying our model (e.g. modifying our simple choice of $p_c$ in Eq.~(3) main text) so that the decay in $p_{\textrm{new}}$ and the
transition to the second scaling occurs already for shorter $\tilde{M}$. For even larger $\tilde{M}$ we do not have data for our model due to
computational limitations. However, based on our asymptotic calculations in Sec.~III, main text, we expect that the observed agreement
will extend over the entire range of available data.

\subsection{Zipf's plot}\label{sec.rwsim.z}

In the analysis of the Zipf's plot~$F(r)$ obtained by our model it is important to take into account that Yule's type processes (already used
words are drawn proportional to their previous occurrences) give a disproportionally large weight to the first word types used in the
simulation. This happens because in the beginning of the simulation there are only a few existing word types into which all drawn tokens are
attributed. Fig.~\ref{fig.SimRW_cut}(a) illustrates this effect and shows that it is inversely proportional to $p_{\mathrm{new}}^0$, which
sets the time-scale for the appearance of new word types in the beginning of the simulation. This artifact can be easily addressed by
excluding a few word types of smallest rank and re-normalizing the remaining distribution, as shown in
Fig.~\ref{fig.SimRW_cut}(b-d). Alternatively, one can modify the 
preferential attachment part of the model (right-most branch in Fig. 3, main text) so that the very first used word-types follow a different
rule and have a vanishing probability of usage for large $M$ (notice that this would not modify the Heap's plot).  For the case of English,
Fig.~\ref{fig.SimRW_z} shows that the removal of only one word type ($r=1$) is sufficient in order to obtain a good agreement with data
(less than $50\%$ of deviation over $7$ orders of magnitude). As discussed in the case of Heaps' plot above, the transition to the second scaling appears slightly
shifted in comparison to the data.


\begin{thebibliography}{55}
\bibitem {footnote:dp}%
  \BibitemOpen
The sharp transition between the two regimes in Eq.~(\ref{eq.modeldp}) might seem artificial. We believe that alternative
  distributions which interpolate between the two scalings could provide a similarly good account of the data. The advantage of the
  distribution Eq.~(\ref{eq.modeldp}) is that the transition point $r=b$ appears explicitely as a free parameter and can be independently
  estimated from data
  \BibitemShut {NoStop}%
\bibitem {footnote:ex}%
  \BibitemOpen
These examples are the $6$ most frequent words which belonged to the core vocabulary (i.e., $r<b^*=7873$) in every single year in one decade
and in none of the years in the other decade (ordered by the average frequency in the decade which they belonged to the core vocabulary)
  \BibitemShut {NoStop}%
\end{thebibliography}
%
\newpage
\clearpage
%
%
\begin{figure*}
\centering
\includegraphics[width=1\columnwidth]{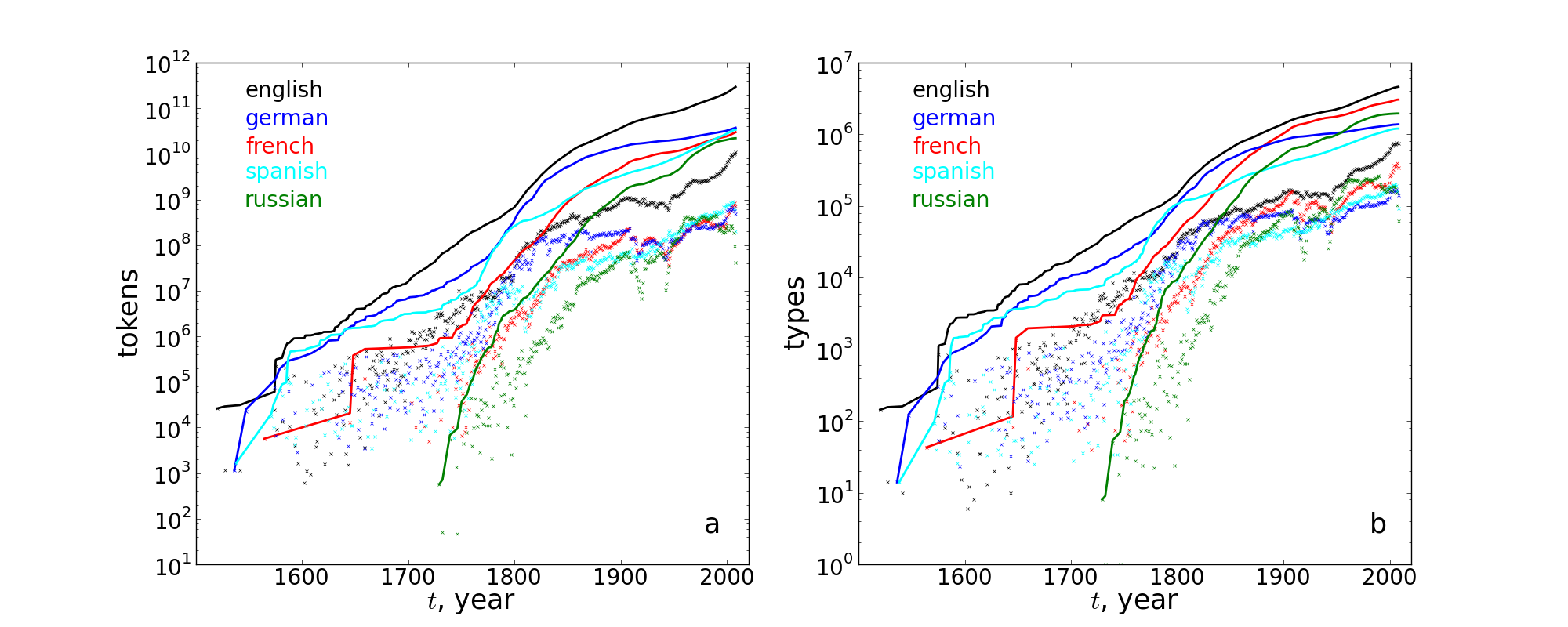}
\caption{Size of the database after filtering. a) Number of tokens for yearly data $y(t)$ (x-symbols) and cumulative data $Y(t)$ (line). b) Number of types for yearly data $y(t)$ (x-symbols) and cumulative data $Y(t)$ (line). Each language is marked by a different color.}
\label{fig.Size}
\end{figure*}
%
%
\begin{figure*}
\centering
\includegraphics[width=0.66\columnwidth]{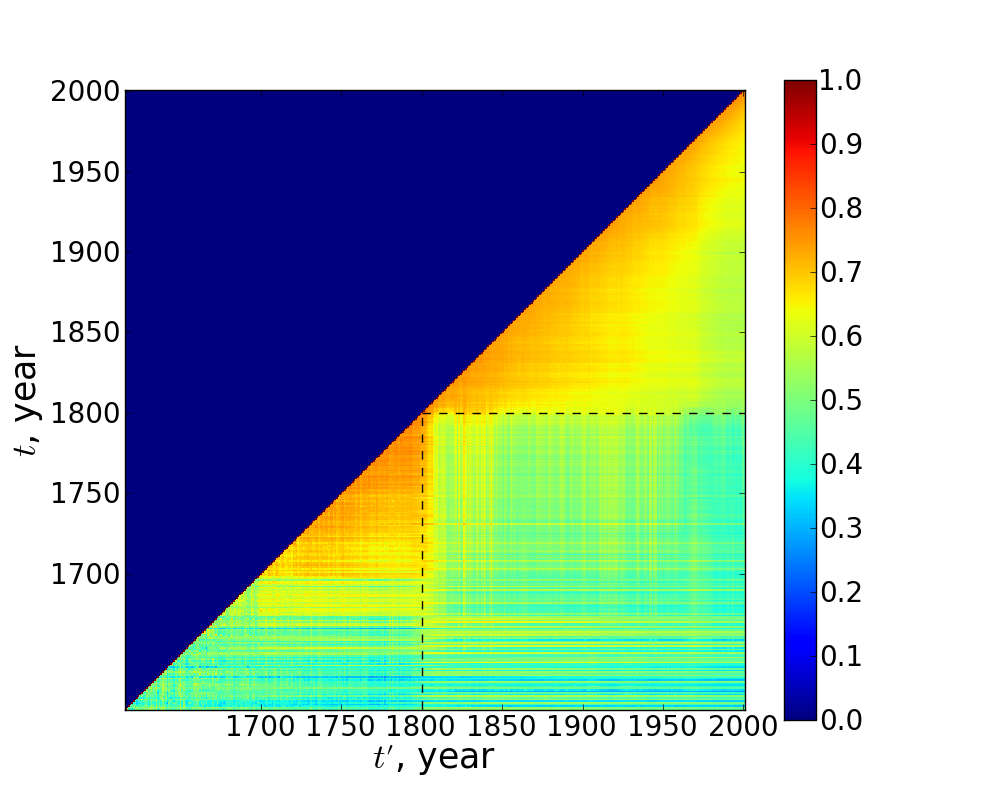}
\caption{Correlation between data in different years for English. Kendall's rank correlation Eq.~(\ref{eq.Kend.tau}) between yearly data $y(t)$, $y(t')$ for $t,t' \in [1500,2000]$ with $t\leq t'$. The dasehd lines show $t=1800$ and $t'=1800$ where a noncontinuous change in the correlation occurs. }
\label{fig.Rank}
\end{figure*}
%
%
\begin{figure*}
\centering
\includegraphics[width=.66\columnwidth]{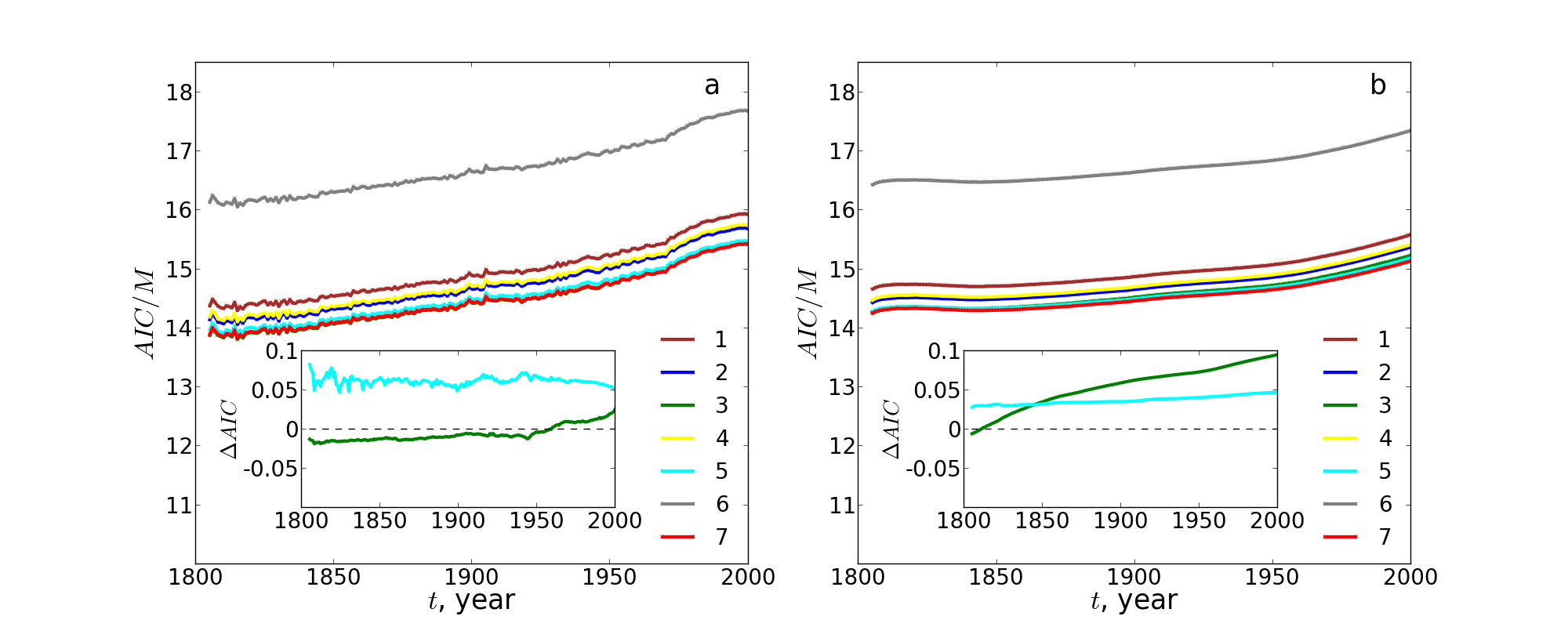}
\includegraphics[width=.33\columnwidth]{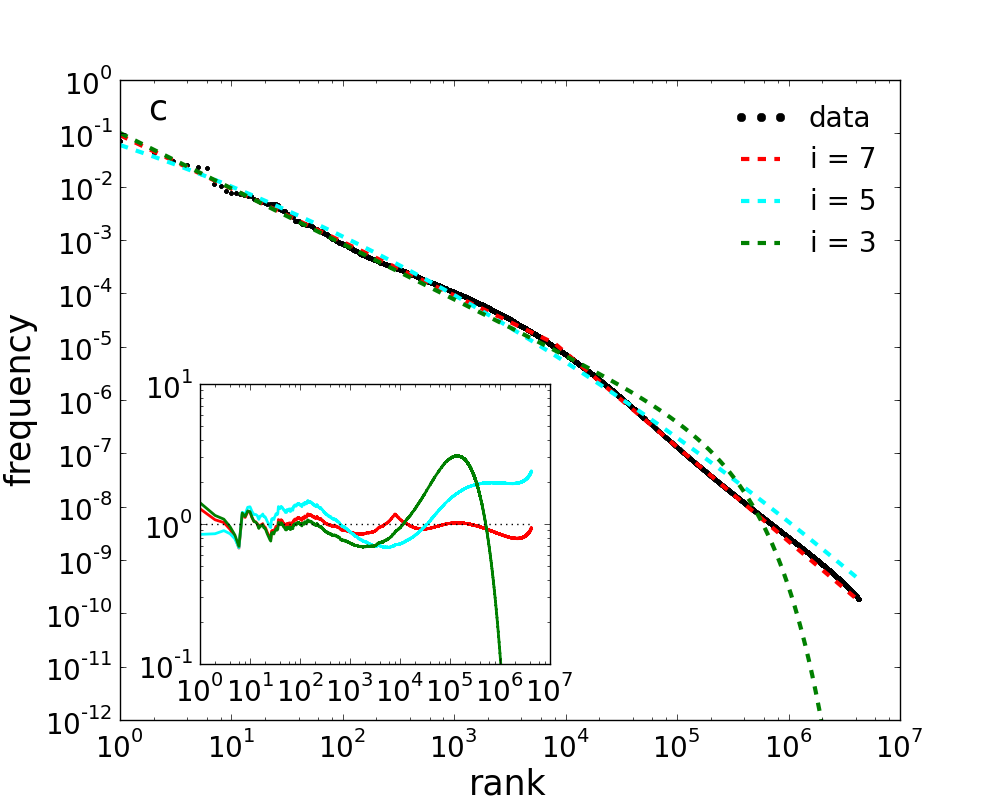}
\caption{Discrimination between different models with $AIC$ for English. Value of the $AIC$ for a) yearly data $y(t)$ b) cumulative data $Y(t)$. The inset shows the difference $\Delta AIC = AIC_{\mathrm{i}}/M - AIC_{7}/M$, $i=1..6$ meaning that if $\Delta AIC >0$ the double power-law is the most likely model among the proposed describing the data. Numbers refer to the indices of the model in Tab.~\ref{tab.distmodels}. c) rank-frequency plot for $Y(2000)$ and the fits of the three best models. The inset shows the ratio $F_{\text{data}}(r)/F_{\text{fit}}(r)$. }
\label{fig.AIC.eng}
\end{figure*}

\begin{figure*}
\centering
\includegraphics[width=0.66\columnwidth]{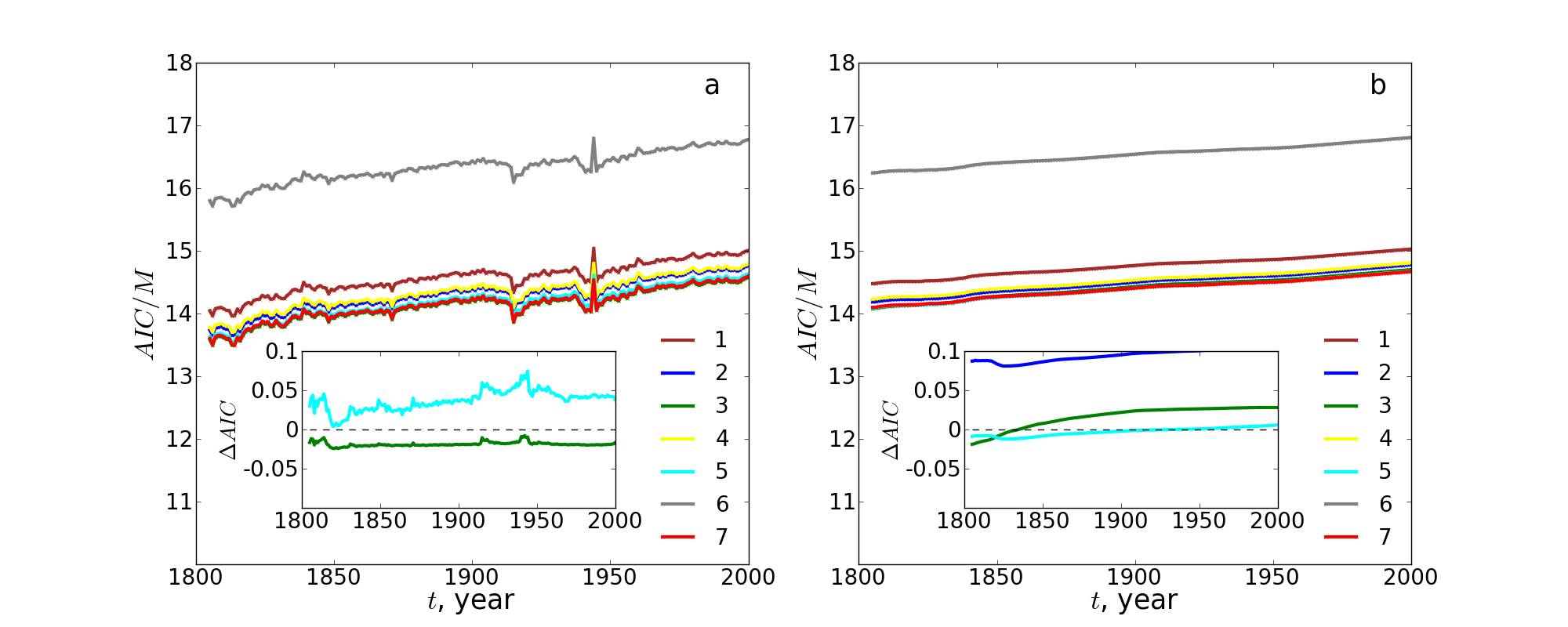}
\includegraphics[width=.33\columnwidth]{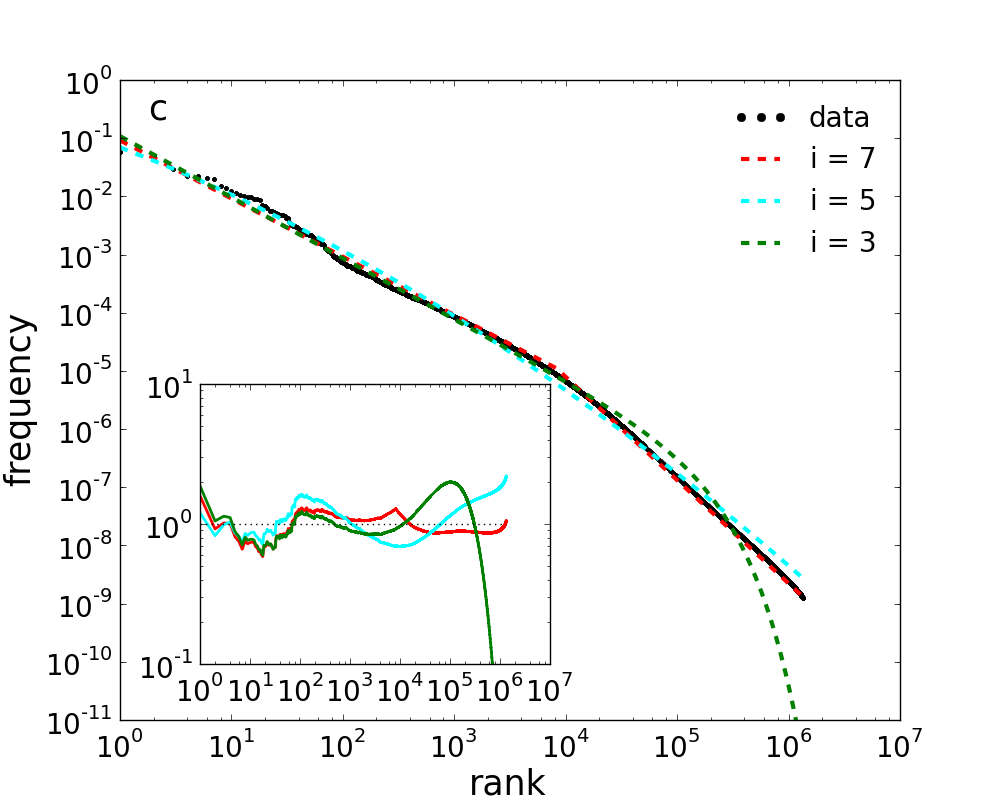}
\caption{Same as in Fig.~\ref{fig.AIC.eng} for French.  }
\label{fig.AIC.fre}
\end{figure*}

\begin{figure*}
\centering
\includegraphics[width=0.66\columnwidth]{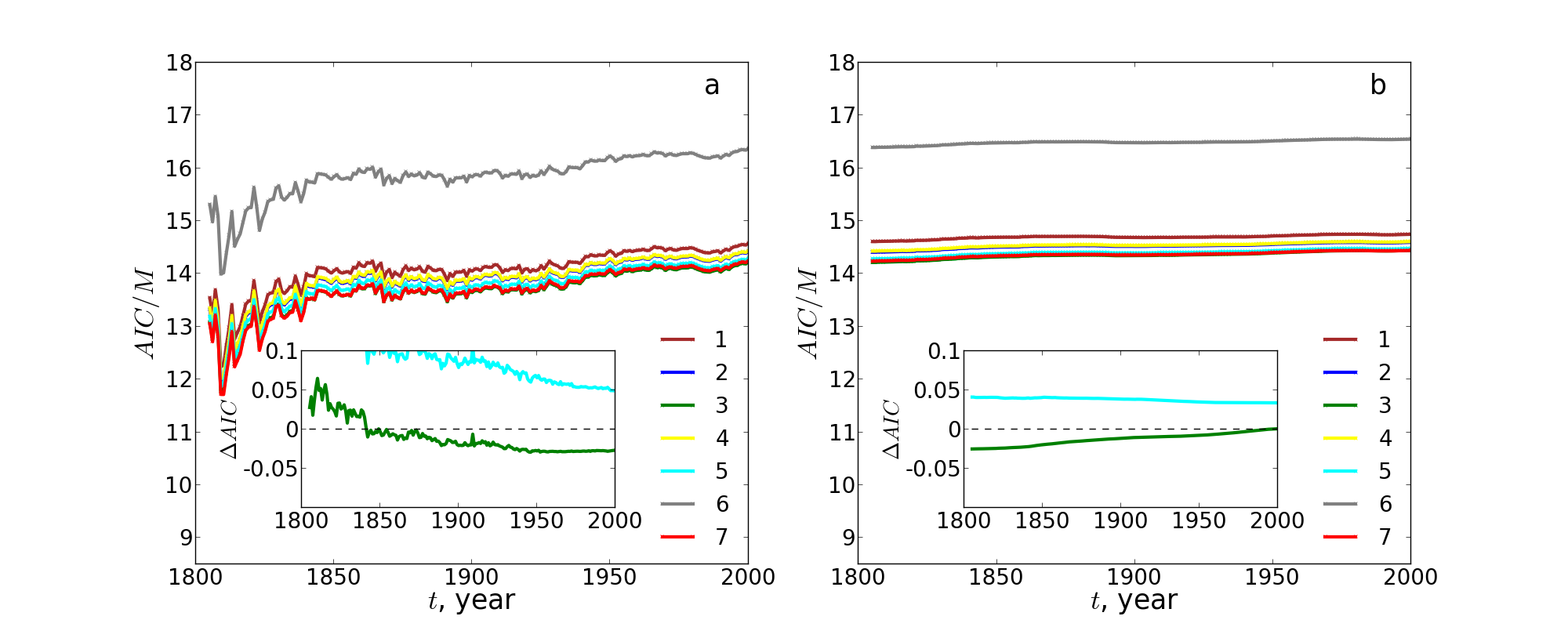}
\includegraphics[width=.33\columnwidth]{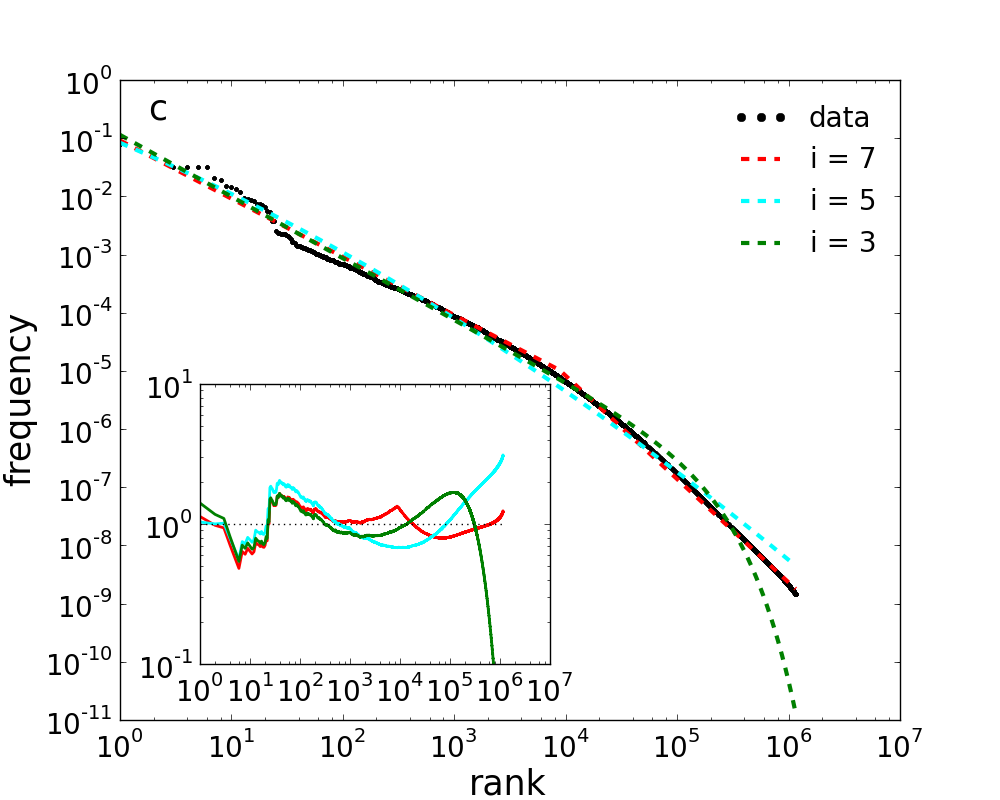}
\caption{Same as in Fig.~\ref{fig.AIC.eng} for Spanish.  }
\label{fig.AIC.spa}
\end{figure*}

\begin{figure*}
\centering
\includegraphics[width=0.66\columnwidth]{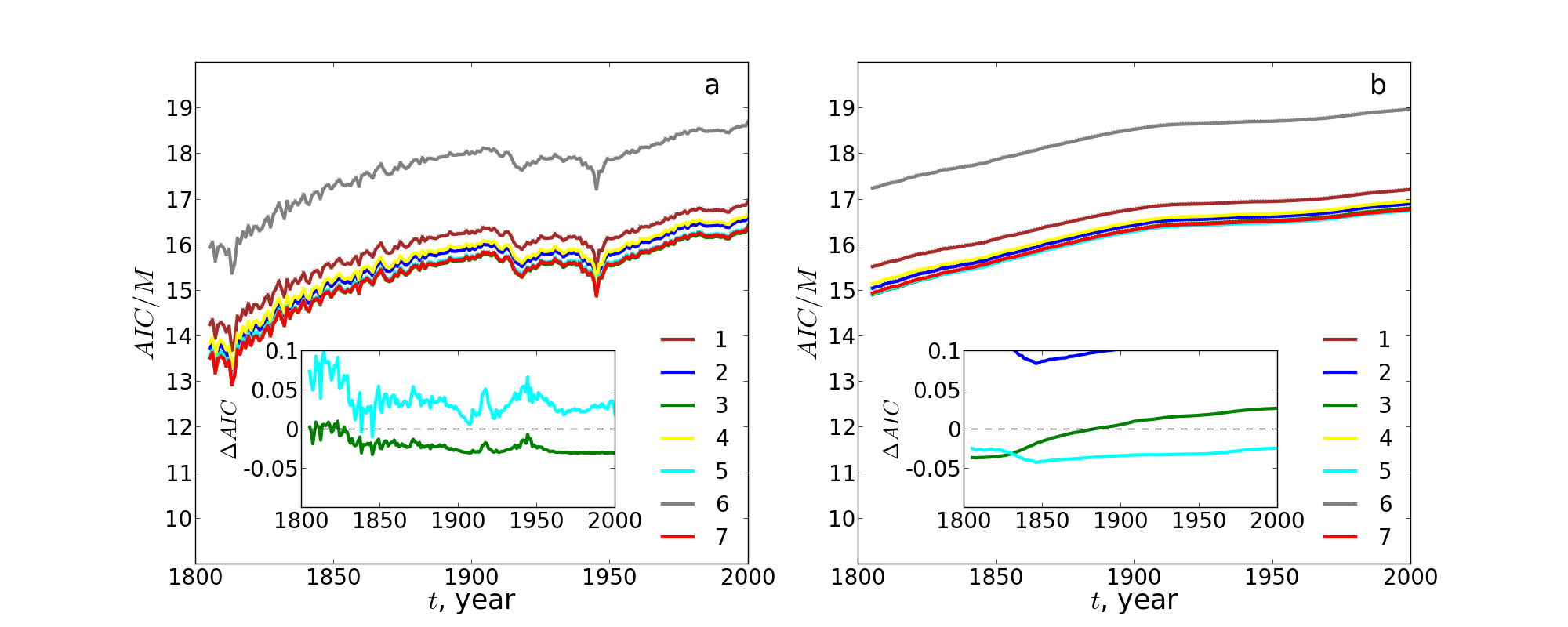}
\includegraphics[width=.33\columnwidth]{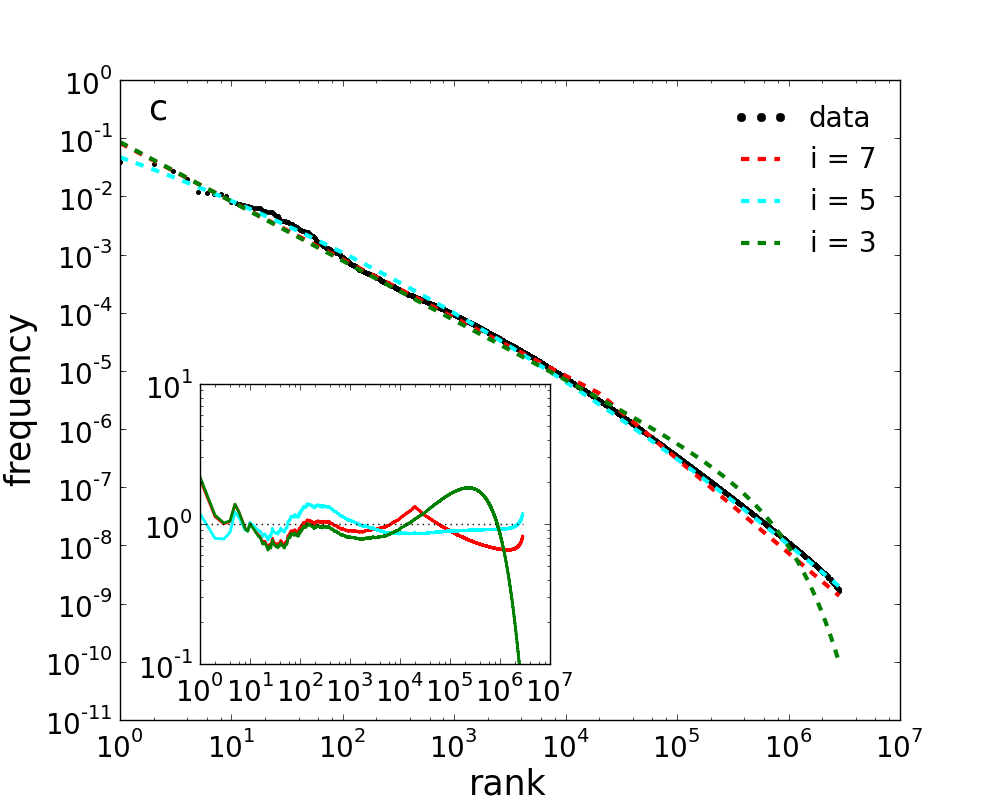}
\caption{Same as in Fig.~\ref{fig.AIC.eng} for German.  }
\label{fig.AIC.ger}
\end{figure*}

\begin{figure*}
\centering
\includegraphics[width=0.66\columnwidth]{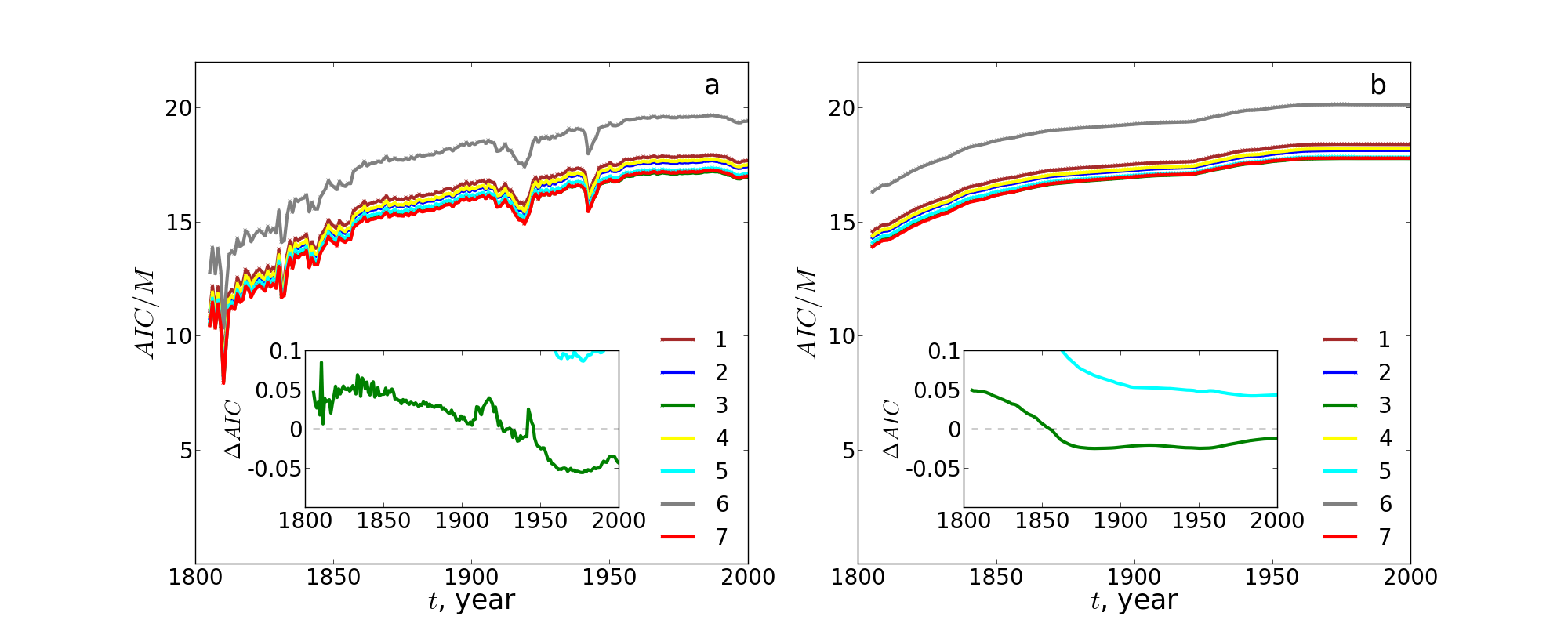}
\includegraphics[width=.33\columnwidth]{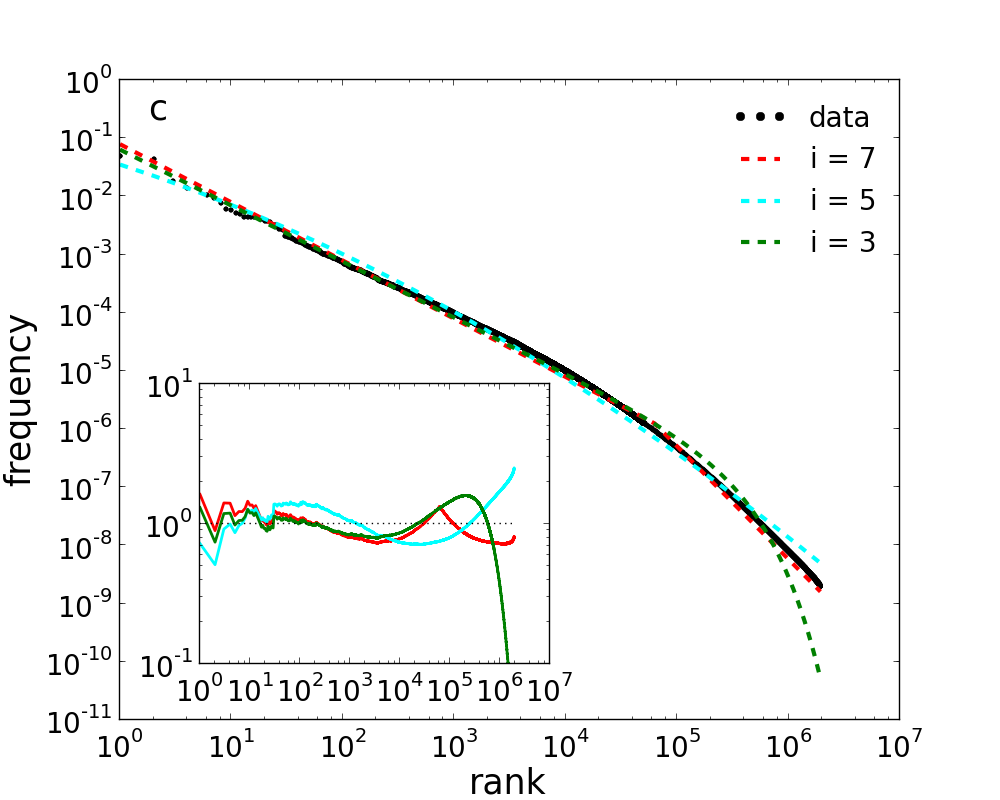}
\caption{Same as in Fig.~\ref{fig.AIC.eng} for Russian.  }
\label{fig.AIC.rus}
\end{figure*}

\begin{figure*}
\centering
\includegraphics[width=.66\columnwidth]{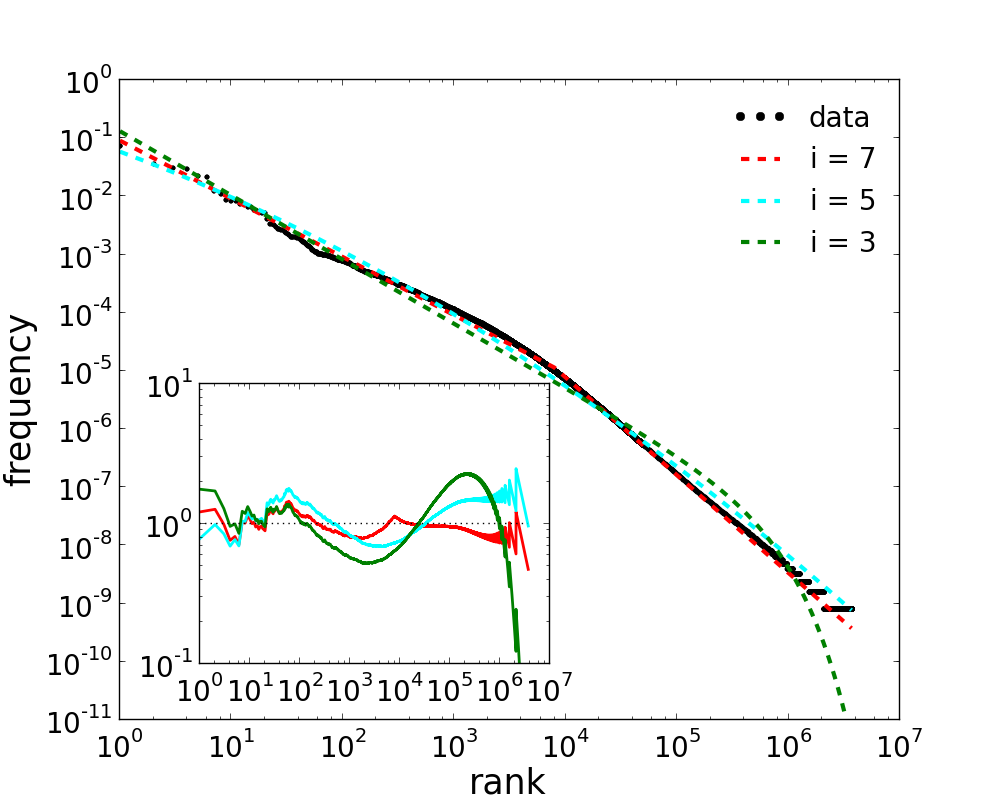}
\caption{Rank frequency distribution for the English Wikipedia and the fits of the three best models. The inset shows the ratio $F_{\text{data}}(r)/F_{\text{fit}}(r)$.   }
\label{fig.AIC.wikien}
\end{figure*}

%
%
\begin{figure*}
\centering
\includegraphics[width=0.66\columnwidth]{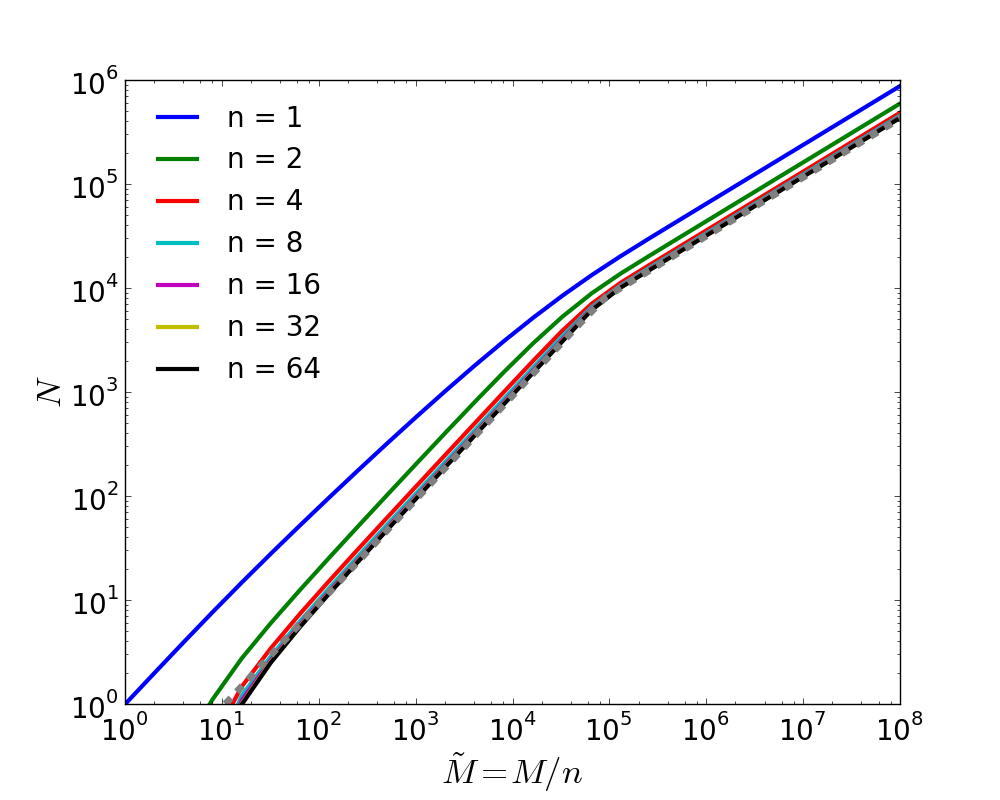}
\caption{Influence of threshold $n$ on size of vocabulary for the ZE. Growth curves $N(\tilde{M}=M/n)$ obtained from ZE for double power-law (Eq.(1) main text) with parameters $\gamma^*=1.77$, $b^*=7873$  with different thresholds $n$. The dashed curve shows the asymptotic solution Eq.~(\ref{eq.ZEasym}). }
\label{fig.Heapskmin.ze}
\end{figure*}

\begin{figure*}
\centering
\includegraphics[width=1.0\columnwidth]{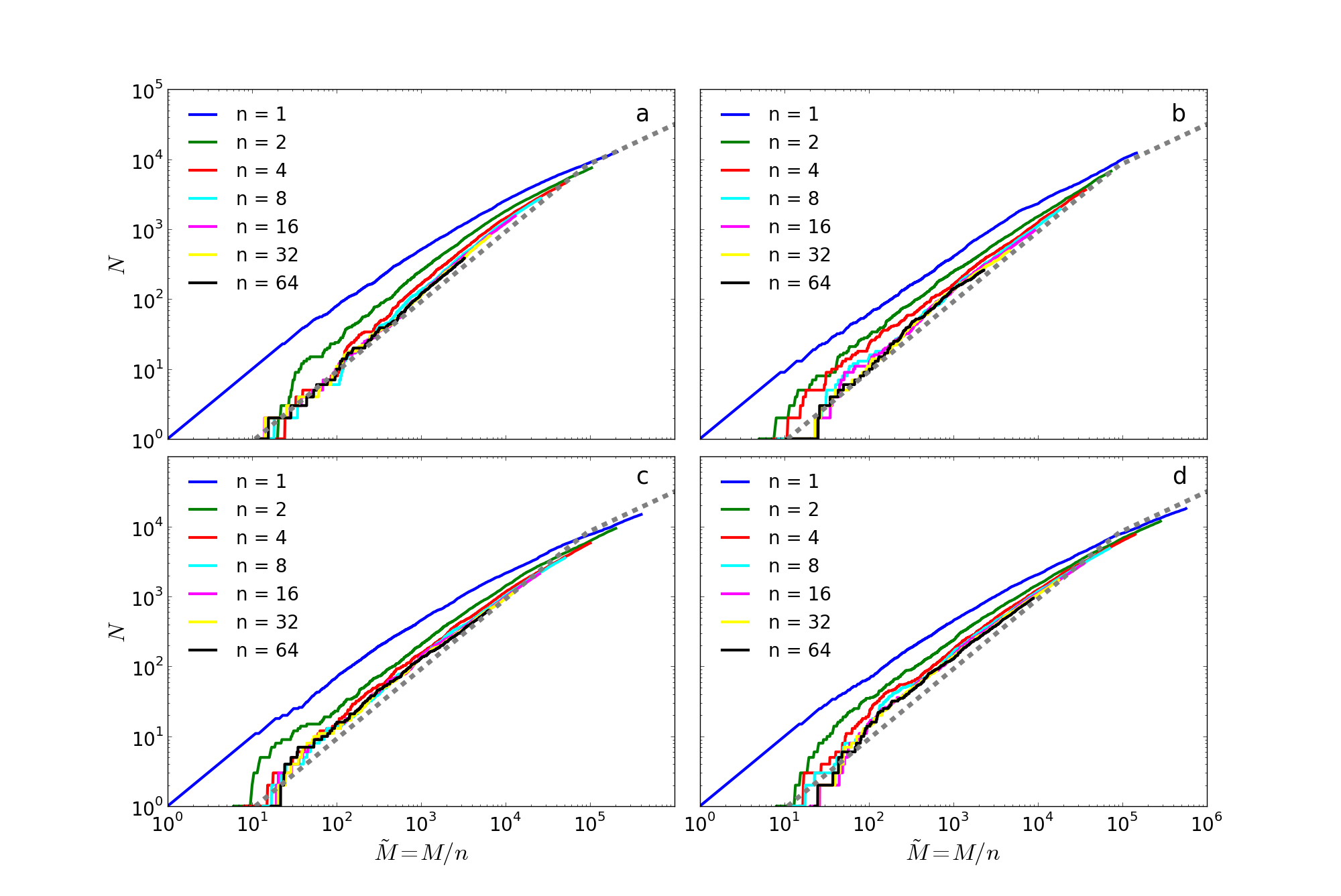}
\caption{Influence of threshold $n$ on size of vocabulary for single books. Growth curves $N(\tilde{M}=M/n)$ obtained from 4 different books with different thresholds $n$. a) Charles Darwin: ``The Voyage of the Beagle'' b) Mark Twain: ``Life on the Mississippi''  c) Miguel de Cervantes Saavedra: ``Don Quixote'', translated by John Ormsby d) Leo Tolstoy: ``War and Peace'', translated by Louise and Aylmer Maude. All texts were retrieved from the Project Gutenberg (www.gutenberg.org) on 21.09.2010. The dashed curve shows the asymptotic solution Eq.~(\ref{eq.ZEasym}) of the ZE assuming a double power-law (Eq.(1) main text) with parameters $\gamma^*=1.77$, $b^*=7873$.}
\label{fig.Heapskmin.books}
\end{figure*}

\begin{figure*}
\centering
\includegraphics[width=0.66\columnwidth]{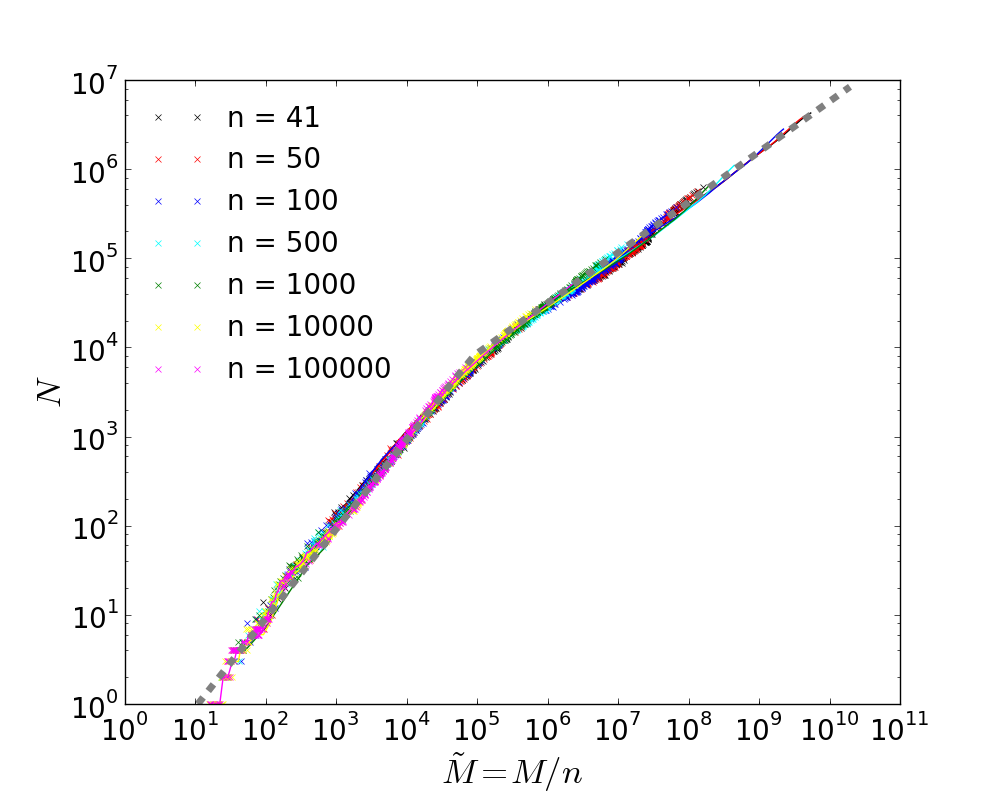}
\caption{Influence of threshold $n$ on size of vocabulary for the English google-ngram database. Growth curves $N(\tilde{M}=M/n)$ obtained from yearly data $y(t)$ (x-symbol) and cumulative data $Y(t)$ (line) for different values of the threshold $n$ with $n\in [41,10^5]$ marked by different colors. The dashed curve shows the asymptotic solution Eq.~(\ref{eq.ZEasym}) of the ZE assuming a double power-law (Eq.(1) main text) with parameters $\gamma^*=1.77$, $b^*=7873$. }
\label{fig.Heapskmin.data}
\end{figure*}
%
%
\begin{figure*}
\centering
\includegraphics[width=0.66\columnwidth]{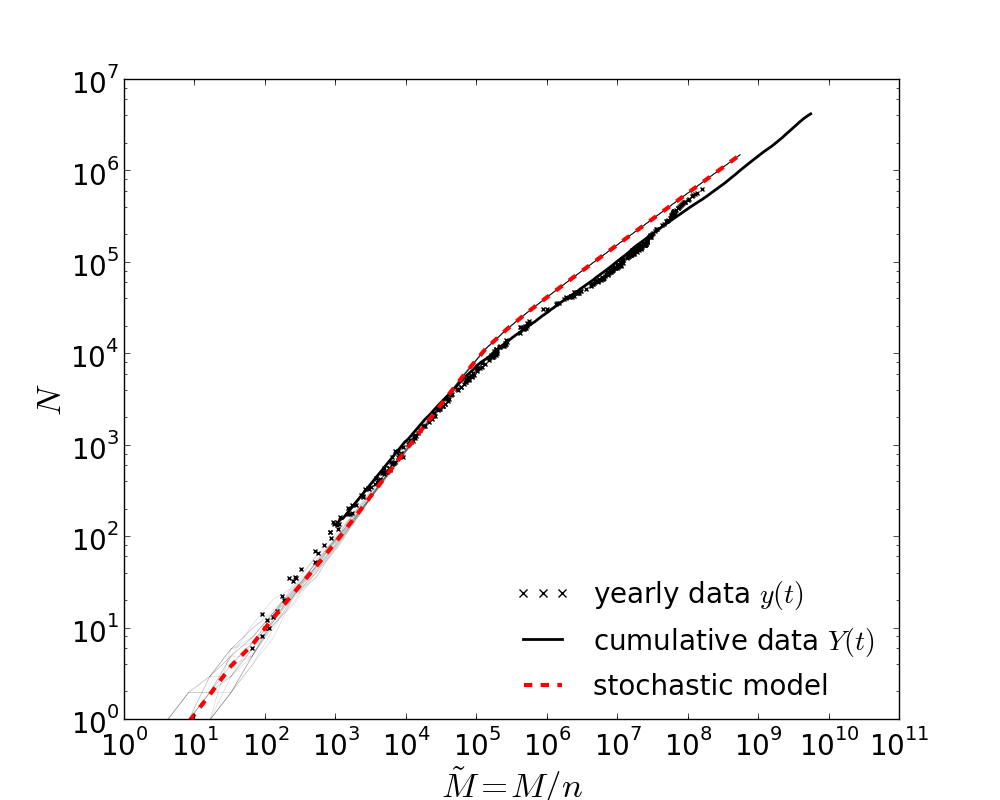}
\caption{Vocabulary growth, $N(\tilde{M})$, from the numerical simualtion of our stochastic model (Heaps' plot). Number of word-types as a function of word-tokens of the English database for yearly (x-symbols) database, cumulative (solid) database, and the expectation from our stochastic model (dashed). Single realizations of the stochastic process are shown in thin/gray (solid). Each realization is calculated for an imaginary text of $\tilde{M}=10^9$ tokens.}
\label{fig.SimRW_h}
\end{figure*}
\begin{figure*}
\centering
\includegraphics[width=0.49\columnwidth]{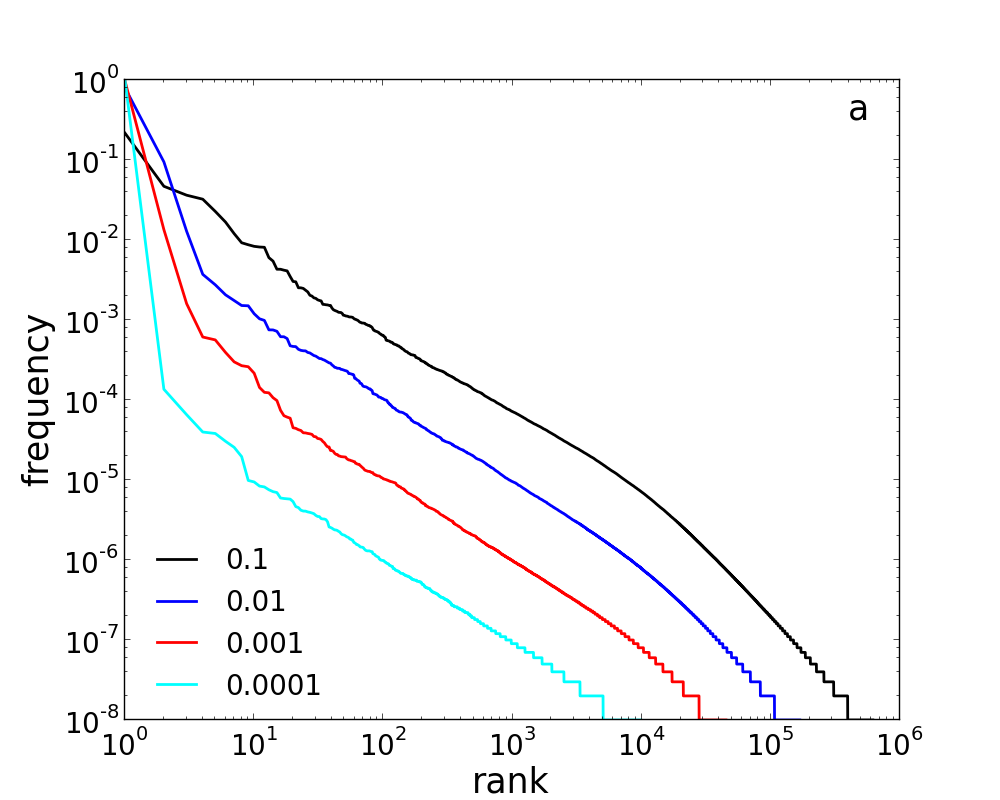}
\includegraphics[width=0.49\columnwidth]{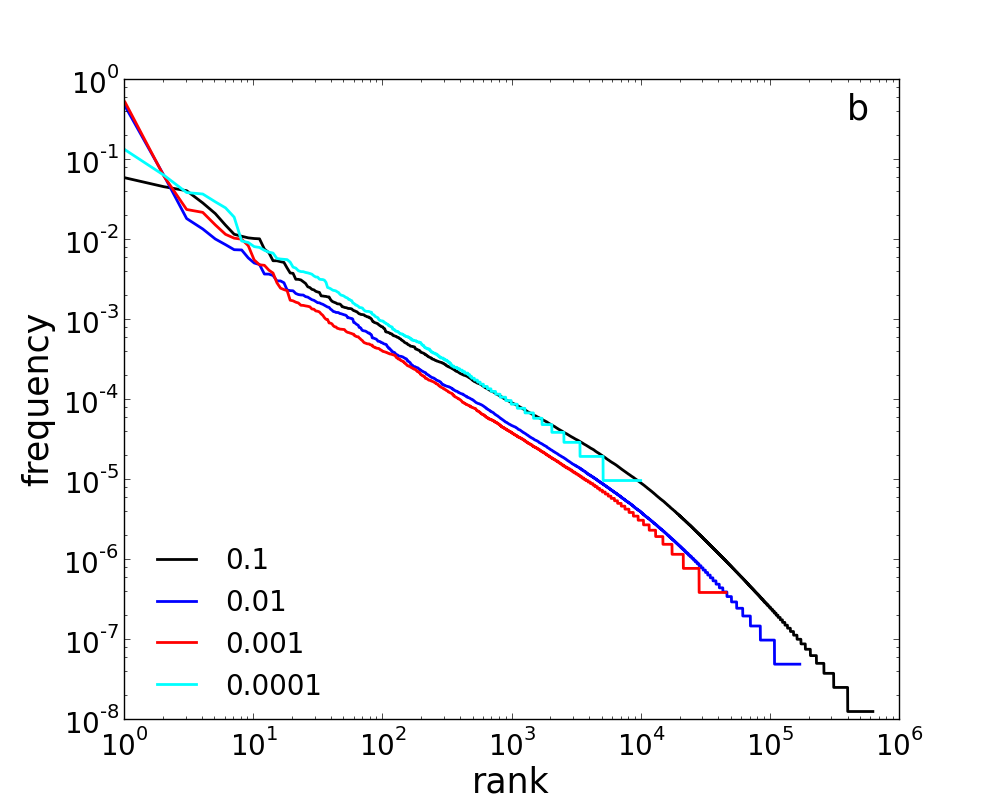}
\includegraphics[width=0.49\columnwidth]{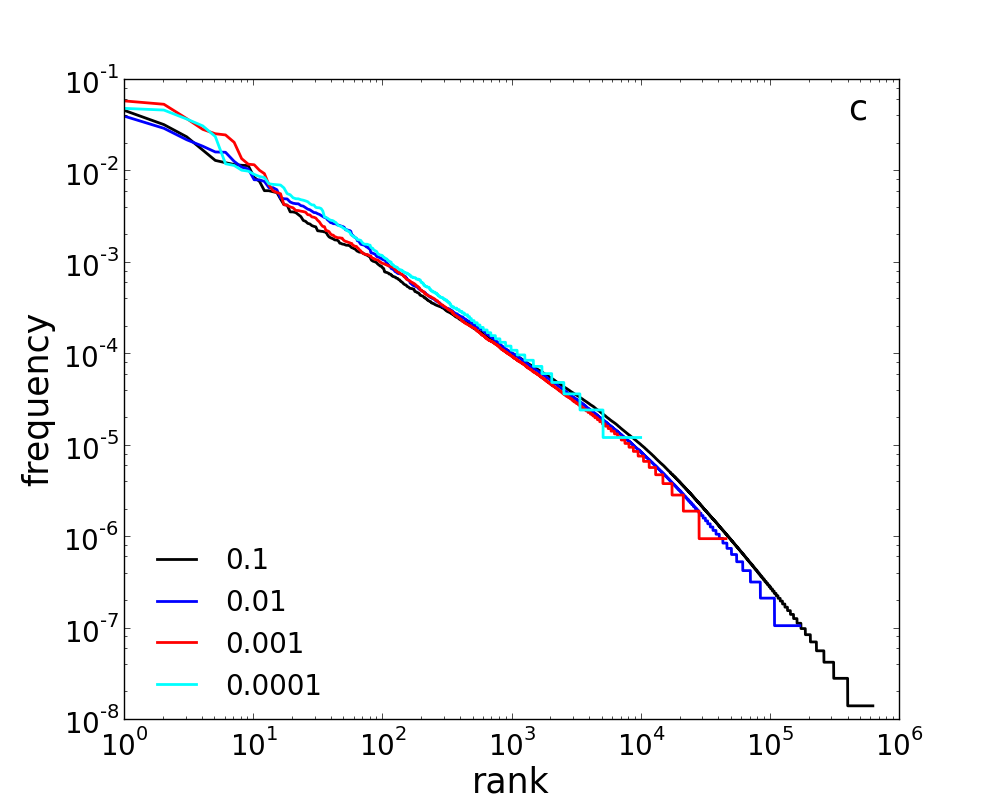}
\includegraphics[width=0.49\columnwidth]{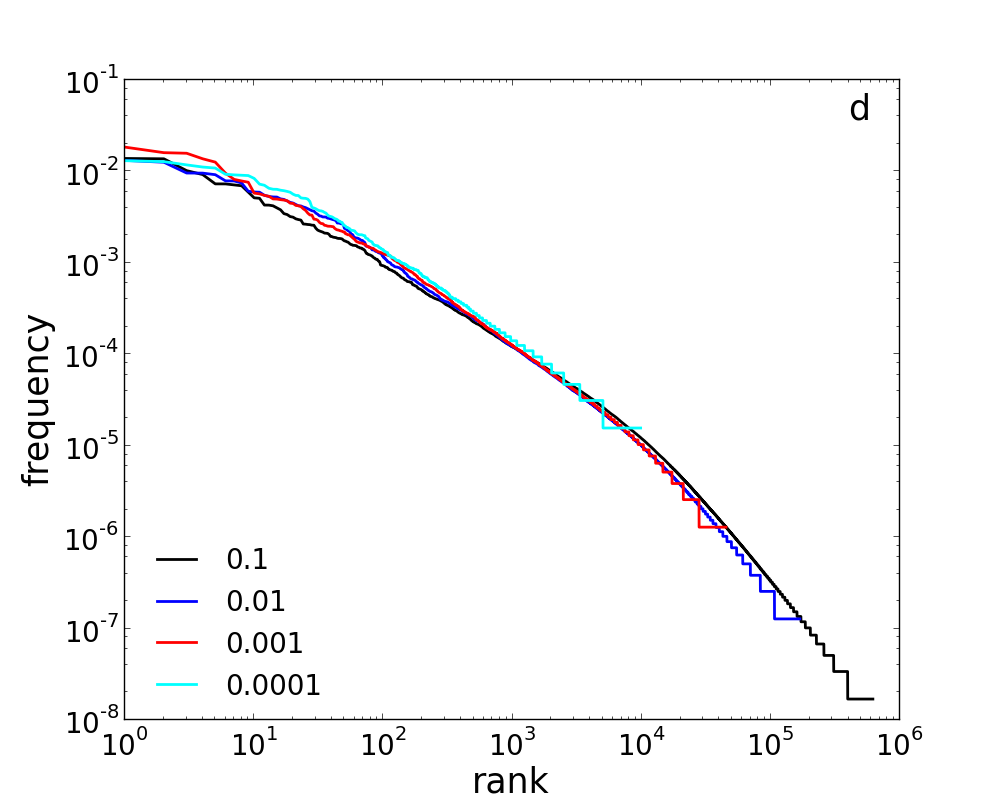}

\caption{Influece of the first word types on the rank-frequency distribution of our model. Rank-frequency distribution $F(r)$ from our numerical simulation with different values for $p_{\mathrm{new}}^0\in \{ 0.1, 0.01, 0.001, 0.0001\}$ after filtering the $k$ most frequent types, where a) $k=0$, b) $k=1$, c) $k=3$, and d) $k=10$. In this context, filtering means, that i) we neglect all tokens associated with ranks $r=1...k$; ii) the rank of all remaining types is lowered by $k$, e.g., the rank of the $k+1$-th most frequent type becomes $r=1$; and iii) the distribution is renormalized such that $\sum_{r=1}^{N-k}F(r)=1$, where $N$ is the number of types before the filtering.    }
\label{fig.SimRW_cut}
\end{figure*}
\begin{figure*}
\centering
\includegraphics[width=0.49\columnwidth]{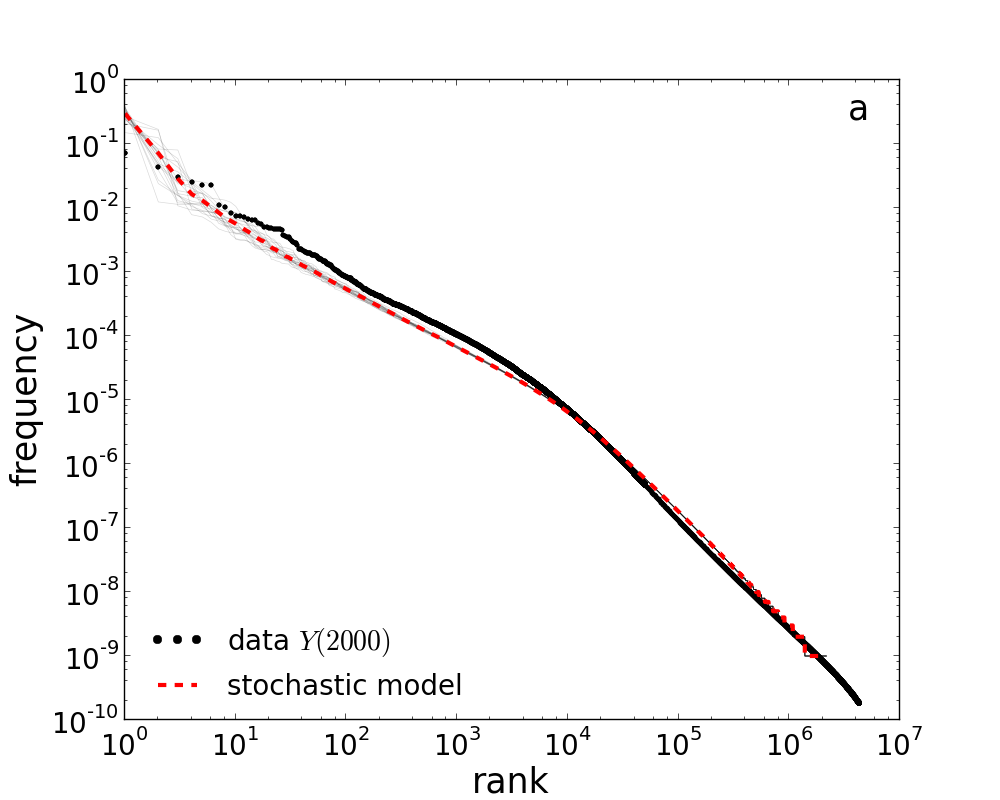}
\includegraphics[width=0.49\columnwidth]{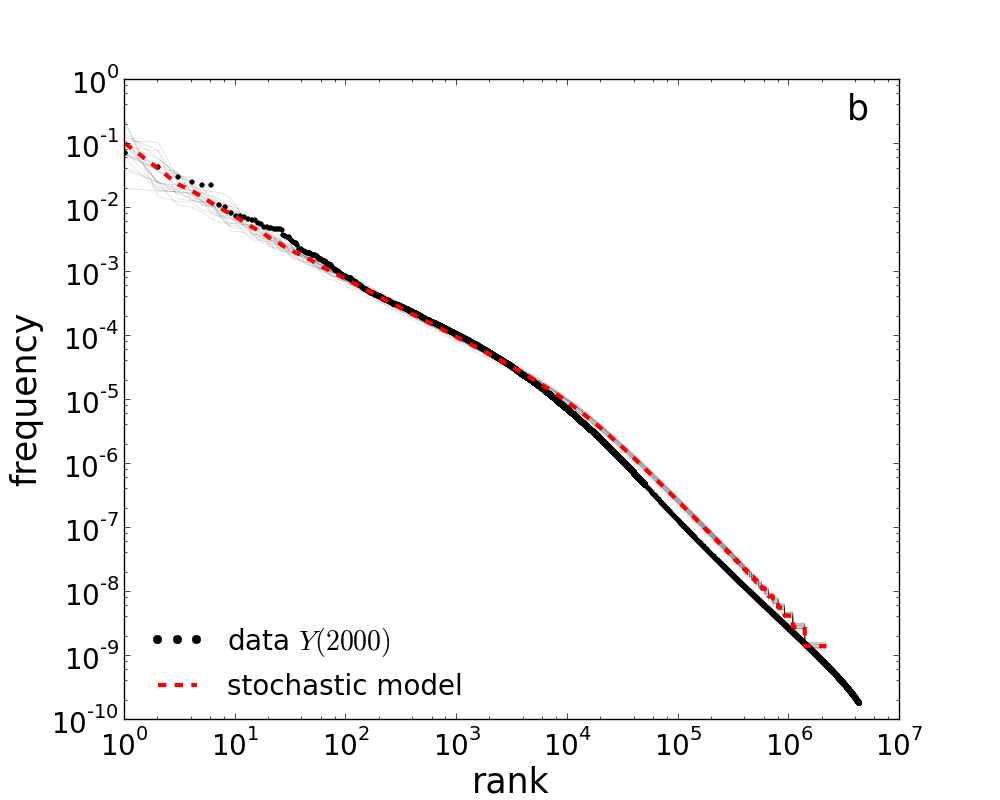}
\caption{Rank frequency distribution, $F(r)$, from the numerical simualtion of our stochastic model (Zipf's plot). Rank-frequency distribution for the English database $Y(2000)$ (solid) and the expectation from our stochastic model (dashed), where a) shows the unfiltered result, and b) shows the distribution after filtering the type which has rank $r=1$ in a). Single realizations of the stochastic process are shown in thin/gray (solid). Each realization is calculated for an imaginary text of $\tilde{M}=10^9$ tokens.}
\label{fig.SimRW_z}
\end{figure*}

\end{widetext}
\end{document}